\pdfoutput=1
\documentclass[sigconf]{acmart}
\renewcommand\footnotetextcopyrightpermission[1]{} 
\def\BibTeX{{\rm B\kern-.05em{\sc i\kern-.025em b}\kern-.08emT\kern-.1667em\lower.7ex\hbox{E}\kern-.125emX}}

\usepackage{adjustbox}
\usepackage{graphicx,amssymb,lineno}
\usepackage{epstopdf}
\graphicspath{{./}{png/}{../}}
\usepackage{algorithm}
\usepackage{algorithmic}
\usepackage{multirow}
\usepackage[tight,footnotesize]{subfigure}
\usepackage{footmisc}
\usepackage{footnpag}
\usepackage{amsmath,amsfonts}
\usepackage{color}
\usepackage{perpage}
\usepackage{fixltx2e}
\usepackage{verbatim}
\usepackage{float}
\usepackage{braket}
\usepackage{nth}
\usepackage{multirow}
\usepackage{epsfig}
\usepackage{textcomp}
\usepackage{xcolor}
\usepackage{makecell}

\setcopyright{acmcopyright}
\copyrightyear{2019} 
\acmYear{2019} 
\acmConference[SC '19]{The International Conference for High Performance Computing, Networking, Storage, and Analysis}{November 17--22, 2019}{Denver, CO, USA}
\acmBooktitle{The International Conference for High Performance Computing, Networking, Storage, and Analysis (SC '19), November 17--22, 2019, Denver, CO, USA}
\acmPrice{15.00}
\acmDOI{10.1145/3295500.3356155}
\acmISBN{978-1-4503-6229-0/19/11}

\begin{document}

%
\title{Full-State Quantum Circuit Simulation \\by Using Data Compression}

%
\author{Xin-Chuan Wu}
\affiliation{%
  \department{Department of Computer Science}
  \institution{University of Chicago}
  \city{Chicago}
  \state{Illinois}
}
\email{xinchuan@uchicago.edu}

\author{Sheng Di}
\authornote{Corresponding author: Sheng Di, Mathematics and Computer Science Division, Argonne National Laboratory, Lemont, IL 60439.}
\affiliation{%
  \department{Mathematics and Computer Science Division}
  \institution{Argonne National Laboratory}
  \city{Lemont}
  \state{Illinois}
}
\email{sdi1@anl.gov}

\author{Emma Maitreyee Dasgupta}
\affiliation{%
  \department{Department of Computer Science}
  \institution{University of Chicago}
  \city{Chicago}
  \state{Illinois}
}
\email{edasgupta@uchicago.edu}

\author{Franck Cappello}
\affiliation{%
  \department{Mathematics and Computer Science Division}
  \institution{Argonne National Laboratory}
  \city{Lemont}
  \state{Illinois}
}
\email{cappello@mcs.anl.gov}

\author{Hal Finkel}
\affiliation{%
  \department{Argonne Leadership Computing Facility}
  \institution{Argonne National Laboratory}
  \city{Lemont}
  \state{Illinois}
}
\email{hfinkel@anl.gov}

\author{Yuri Alexeev}
\affiliation{%
  \department{Computational Science Division}
  \institution{Argonne National Laboratory}
  \city{Lemont}
  \state{Illinois}
}
\email{yuri@anl.gov}

\author{Frederic T. Chong}
\affiliation{%
  \department{Department of Computer Science}
  \institution{University of Chicago}
  \city{Chicago}
  \state{Illinois}
}
\email{chong@cs.uchicago.edu}

\ccsdesc[500]{Computer systems organization~Quantum computing}
%

\renewcommand{\shortauthors}{X. Wu et al.}
\renewcommand{\shorttitle}{Full-State Quantum Circuit Simulation by Using Data Compression}
%
\begin{abstract}
Quantum circuit simulations are critical for evaluating quantum algorithms and machines. However, the number of state amplitudes required for full simulation increases exponentially with the number of qubits. In this study, we leverage data compression to reduce memory requirements, trading computation time and fidelity for memory space. Specifically, we develop a hybrid solution by combining the lossless compression and our tailored lossy compression method with adaptive error bounds at each timestep of the simulation. Our approach optimizes for compression speed and makes sure that errors due to lossy compression are uncorrelated, an important property for comparing simulation output with physical machines.
Experiments show that our approach reduces the memory requirement of simulating the 61-qubit Grover's search algorithm from 32 exabytes to 768 terabytes of memory on  Argonne's Theta supercomputer using 4,096 nodes. The results suggest that our techniques can increase the simulation size by 2$\sim$16 qubits for general quantum circuits. 
\end{abstract}
%
%
%
%
\maketitle

\section{Introduction}\label{sec:intro}
Classical simulation of quantum circuits is crucial for better understanding the operations and behaviors of quantum computation. Such simulations allow researchers and developers to evaluate the complexity of new quantum algorithms and validate quantum devices. The path toward building Noisy Intermediate-Scale Quantum (NISQ) \cite{preskill2018quantum} machines such as IBM's 50-qubit quantum computer and Google's 72-qubit quantum computer \cite{kelly2018preview} will require intermediate-scale quantum circuit simulators to calibrate and verify the hardware.
\begin{table}
\begin{center}
  \caption{\textbf{Examples of supercomputers, their total memory capacity, and the maximum number of qubits they can simulate for arbitrary circuits.}}
  \vspace{-2mm}
  \begin{tabular}{ l  c  c }
    \hline
    System & Memory (PB) & Max Qubits\\ \hline\hline
    Summit & 2.8 & 47\\ \hline
    Sierra & 1.38 & 46\\ \hline
    Sunway TaihuLight & 1.31 & 46 \\\hline
    Theta & 0.8 & 45\\ 
    \hline
  \end{tabular}
\label{tab:supercomputers}
\end{center}
\end{table}

Unfortunately, today's practical full-simulation limit is 47 qubits (Table \ref{tab:supercomputers}) \cite{strohmaier2018top500}. The reason is that the number of quantum state amplitudes required for the full simulation increases exponentially with the number of qubits, making physical memory the limiting factor. Given $n$ quantum bits (qubits), we need $2^n$ amplitudes to describe the quantum system \cite{nielsen2002quantum}. In order to describe the amplitudes precisely, double-precision complex numbers are used to represent the state of the quantum systems. As a result, the size of the quantum state in the simulation is $2^{n+4}$ bytes. Although 49-qubit simulations will become possible in the near future with the arrival of exascale supercomputers \cite{aurora2021},   a gap still remains between the size of classical simulation and the size of NISQ machines. 

Several simulation techniques related to Feynman paths \cite{bernstein1997quantum} and tensor network contractions \cite{boixo2017simulation, markov2008simulating, pednault2017breaking} have been proposed to trade time for space complexity. Different approaches have different benefits and disadvantages. For Feynman paths method, both time and space complexity grow exponentially with the circuit depth, and thus this technique can simulate only shallow circuits. For tensor network simulation methods, since the time complexity grows exponentially with the underlying graph treewidth, these simulation techniques can only simulate the circuits with low treewidth. Some approaches calculate only a single amplitude or a partial state vector in order to be less restrictive on computational resources \cite{chen2018classical, chen201864, pednault2017breaking, boixo2017simulation}. As for quantum software development, several types of quantum applications require intermediate measurement \cite{PhysRevA.54.3824, gottesman1997stabilizer, barends2014superconducting, brassard1996teleportation}. In addition, recent studies focus on quantum software debugging by inserting assertions in the middle of quantum programs \cite{huang2019stat}. Tensor network simulation techniques do not effectively support intermediate measurement and full-state assertion checking for software debugging. At the same time, NISQ machines are evolving toward supporting deeper circuits with error mitigation techniques \cite{bonet2018low, kandala2019error, endo2018practical, sagastizabal2019error}, which may make full-state simulations of high-depth circuits more important. When trying to verify quantum hardware and software by using a full-state simulator, every qubit counts in maximum simulation size.  Every qubit closer to physical machine sizes means 2X more state space that can be evaluated, or, conversely, 2X smaller gap in state space between the simulation and the physical hardware.

To simulate general circuits with higher qubit count and depth, we propose quantum circuit simulation techniques that can reduce the memory requirement of the full simulation by compressing quantum state amplitudes at runtime. Specifically, we apply data compression techniques to the quantum state vector during the simulation. Since we aim to simulate intermediate-scale general quantum circuits, we have to achieve a data compression ratio as high as possible, because the compression ratio is the key to increasing the number of qubits in the simulation. Our approach uses lossless compression, lossy compression, and adaptive error bounds to reduce the memory requirement of the simulation. In general, lossy compression algorithms lead to significantly higher compression ratios than do lossless compressors, while introducing errors to a certain extent. To minimize the error propagation and guarantee high-fidelity simulation results, we utilize both Zstandard lossless compressor \cite{zstd} and an error-bounded tailored lossy compressor in our simulation framework. 

Intuitively, one might be concerned that lossy compression would introduce correlated errors in our simulation output that would be very different from the kinds of errors that physical machines would experience.  We will see, however, that our lossy compression is applied in an uncorrelated fashion.  We shall also see that this has an added benefit of dramatically speeding up compression time.

Using our techniques, we are able to trade computation time and fidelity for memory space. We implement our techniques on Intel-QS, a full-state quantum circuit simulator developed by Intel \cite{smelyanskiy2016qhipster}. Intel-QS is an MPI-based distributed high-performance quantum circuit simulator that can run on supercomputing systems. By orchestrating data compression techniques and full-state high-performance simulation techniques, our approach is capable of simulating intermediate-scale general quantum applications and hence obtain effective results of calibration, verification, and benchmarking for NISQ quantum machines.

Our approach integrates knowledge of quantum computation and data compression techniques to reduce the memory requirement of quantum circuit simulations such that our technique allows us to simulate a larger  quantum system with the same memory capacity. We provide one more option in the set of tools to scale quantum circuit simulation. The main contributions of our work are as follows.

\begin{itemize}
\item We present a new technique to reduce memory requirements of full-state simulations of general quantum circuits by using data compression. Reducing memory requirements allows us to increase the number of qubits in our full-state simulations.
\item We design a novel lossy compression method to optimize both compression ratios and compression speed for quantum circuit simulations. This lossy compression technique can be combined with several existing simulation techniques to further reduce the memory footprint.
\item We implement our general quantum circuit simulation framework on the Theta supercomputer at Argonne National Laboratory (ANL). 
\item Our experimental results show that our approach reduces the memory requirement of simulating the 61-qubit Grover's search quantum circuit from 32 exabytes to 768 terabytes of memory. Based on the state-of-the-art simulation techniques, the results suggest that our technique can increase the simulation size by 2 to 16 qubits for general quantum applications with 0.976 simulation fidelity on average. 
\end{itemize}

Our paper is organized as follows. In Section~\ref{background}, we introduce the basics of quantum computation and simulation as well as the data compression techniques. In Section~\ref{overview}, we present our simulation design. In Section~\ref{sec:compression}, we describe our lossy compression technique, and in Section~\ref{evaluation} we evaluate our approach. We discuss future directions and provide conclusions in Section~\ref{conclusion}.

\section{Background and Related Work} \label{background}
In this section, we provide a brief overview of the quantum computation and  discuss related work on quantum circuit simulations. We then present the relevant background on compression techniques.

\subsection{Principles of Quantum Computation}
A qubit is a two-level quantum system, and the state $\ket{\psi}$ can be expressed as 
\begin{equation}
\label{eq:psi2}
\ket{\psi} = a_0\ket{0} + a_1\ket{1},
\end{equation}
where $a_0$ and $a_1$ are complex amplitudes and $|a_0|^2 + |a_1|^2 = 1$. $\ket{0}$, and $\ket{1}$ are two computational orthonormal basis states. The quantum state can also be represented as follows.
\begin{equation}
\label{eq:sumupto1}
\ket{\psi} = a_0\begin{bmatrix}1\\0\end{bmatrix} + a_1\begin{bmatrix}0\\1\end{bmatrix} = \begin{bmatrix}a_0\\a_1\end{bmatrix}
\end{equation}

More generally, the state of an $n$-qubit quantum system can be represented by using $2^n$ amplitudes, as follows.
\begin{equation}
\label{eq:2nrep}
\ket{\psi} = a_{0...00}\ket{0...00} + a_{0...01}\ket{0...01} + ... + a_{1...11}\ket{1...11}
\end{equation}
The squared magnitudes have to sum up to 1, that is,
\begin{equation}
\label{eq:psi}
\sum_i |a_i|^2 = 1.
\end{equation}

In  quantum computation, quantum gates are applied to the quantum system. All gates are represented in matrix form. General single-qubit gates and two-qubit controlled gates are known to be universal \cite{divincenzo1995two}. Applying a single-qubit gate $U$ to the $k$th qubit can be represented by a unitary transformation
\begin{equation}
\label{eq:psi3}
A = I ^{\otimes n-k-1} \otimes U \otimes I^{\otimes k},
\end{equation}
where $I$ is a $2\times 2$ identity matrix and $U$ is a $2\times 2$ unitary matrix.

\subsection{Quantum Circuit Simulation}
Previous studies have provided various types of simulators \cite{listqcsim2019}. Generally, in order to simulate a quantum circuit with $n$ qubits and depth $d$,  there are several simulation approaches \cite{aaronson2016complexity, bernstein1997quantum, markov2008simulating}. Different simulation approaches have different purposes and benefits.\\ 

\noindent\textbf{Schr\"{o}dinger algorithm.} This strategy maintains the full-state vector in memory and updates the state vector in every time step \cite{de2007massively, smelyanskiy2016qhipster}. Since the space grows exponentially with the number of qubits, the physical memory limits the simulation size. The time complexity is polynomial with the number of gates (or circuit depth), and hence this approach is capable of simulating arbitrary depth of circuits. This simulation approach can simulate the supremacy circuits of 45 qubits on the Cori II supercomputing system using 0.5 petabytes \cite{haner20170}, and Li et al. further optimize the simulator specifically for the 49-qubit supremacy circuits \cite{li2018quantum}.\\

\noindent\textbf{Feynman paths algorithm.} This approach calculates the amplitude $a_x$ for any $n$-bit string $x \in \{0,1\}^n$ by following all the paths from a final state to the initial state. For the Feynman paths algorithm, this approach requires $O(2^{dn})$ time to perform the simulation \cite{pednault2017breaking}, and hence this simulation technique is  suitable only for low-depth quantum circuits.\\ 

\noindent\textbf{Tensor network contractions.} This approach uses tensor networks to represent quantum circuits \cite{markov2008simulating, biamonte2017tensor, mccaskey2018validating}. The time and space cost for contracting such tensor networks is exponential with the treewidth of the underlying graphs. Therefore, this approach is impractical to simulate large quantum circuits. Several studies have proposed to use tensor network simulation technique to simulate low-depth supremacy circuits \cite{boixo2017simulation, pednault2017breaking, chen2018classical, villalonga2020establishing}. To trade the simulation fidelity for computational resources, the \emph{approximate} simulation is proposed  \cite{markov2018quantum,boixo2017simulation}. The previous studies of approximate simulations target the overall circuit fidelity at 0.005 \cite{markov2018quantum, villalonga2020establishing}, but if we want to use the simulation results to help calibrate and validate the real machines, we might need higher circuit fidelity. As for quantum software development, several types of quantum applications require intermediate measurement \cite{PhysRevA.54.3824, gottesman1997stabilizer, barends2014superconducting, brassard1996teleportation}. In addition, recent studies propose to insert assertions in quantum programs for software debugging by checking the full-state distribution \cite{huang2019stat}. Tensor network simulation techniques do not effectively support intermediate measurement and full-state assertion checking.\\

Other strategies for simulating quantum circuits also exist. The \emph{Gottesman-Knill theorem} \cite{gottesman1998heisenberg} states that circuits  consisting of only Clifford gates can be efficiently simulated in polynomial time, hence there are simulation techniques for the circuits with a restricted gate library \cite{goldberg2017complexity, gottesman1998heisenberg, aaronson2004improved, bravyi2016improved}. In \cite{zulehner2018advanced}, the decision diagram is used to simulate circuits that consist of Clifford+T gates. This approach exploits redundancies in a quantum state to gain more compact representations. Approximate simulation techniques also have been proposed for circuits using only restricted gates \cite{zulehner2019accuracy, zulehner2019matrix, bravyi2016improved}.

\subsection{Data Compression Techniques}
\label{sec:cmpr-tech-background}

With the vast volumes of data being produced by extreme-scale scientific research and applications, various data compression techniques have been developed for years. Basically, scientific researchers mainly adopt two types of compressors: lossless compressors or error-bounded lossy compressors. Lossless compressors usually adopt both variable-length encoding algorithms (such as Huffman encoding \cite{Huffman} and arithmetic encoding \cite{Arithmetic}) and dictionary coders such as LZ77/78 \cite{lz77}. In most cases, however, lossless compressors such as Gzip \cite{gzip}, Zstd \cite{zstd}, and Blosc \cite{blosC} cannot effectively compress scientific data because the ending mantissa bits of floating-point values are  random such that it is  hard to find exactly the same patterns in the data stream. Some studies \cite{di2016fast} show that lossless compressors always suffer from low compression ratios (around 2:1 in most cases), which is far less than enough for today's extreme-scale high-performance computing (HPC) applications. Accordingly, error-bounded lossy compression has been widely treated as the best solution to such a big scientific data issue, because it not only can significantly reduce the data size but also can control the data distortion according to the user's requirements.

Error-bounded lossy compressors may have distinct designs and implementations, so selecting the most appropriate compression technique is critical to our research. All existing error-bounded lossy compressors can be categorized into two models:  data-prediction based  and domain-transform based, which are described below. 

\begin{itemize}
\item \emph{Data-prediction-based compression model}. This model tries to predict each data point as accurately as possible based on its neighborhood in spatial or temporal dimension and then shrinks the data size by some coding algorithm such as data quantization \cite{tao2017significantly} and bit-plane truncation. A typical example compressor is SZ \cite{Xin-bigdata18}, which involves four compression steps: (1) data prediction, (2) linear-scaling quantization, (3) entropy-encoding, and (4) lossless compression. The errors are introduced and controlled at step (2). Other examples include ISABELA \cite{isabela} and FPZIP \cite{fpzip}.
\item \emph{Domain-transform-based compression model}. This model needs to transform all the original data values to another nonorthogonal coefficient domain for decorrelation and then shrink the data size by applying some optimized coding algorithms such as embedded coding \cite{zfp}. A typical example compressor is ZFP \cite{zfp}, which performs the classic texture compression by leveraging three techniques in each $4^d$ block (where $d$ is the number of dimensions): (1) exponent alignment, (2) (non)orthogonal block transform, and (3) embedded coding. The compression errors (or distortion of data) are introduced and controlled only at step (3). The other two techniques are VAPOR \cite{vapor2} and Sasaki et al.'s compressor \cite{sasaki}, which both adopt the Wavelet transform in the domain transform step.
\end{itemize}

All the existing state-of-the-art error-bounded lossy compressors are  designed or assessed mainly for  visualization, so they are not optimized for the requirement of data fidelity and compression quality in the context of quantum computing simulation. In this sense, we first characterize the effectiveness of the existing state-of-the-art lossy compressors on  quantum circuit simulation results and then exploit a fairly efficient compression method beyond the existing lossy compressors. In our study, we investigate two types of error controls, pointwise absolute error bound and pointwise relative error bound, because they have been supported by the existing state-of-the-art lossy compressors well.
\begin{itemize}
    \item Absolute Error Bound (denoted by \emph{e}). With this type of error control, the compression errors (defined as the difference between the original data value and its corresponding decompressed value) of all data points must be strictly limited in the required bound; in other words, $d'_i$ must be in [$d_i-e,d_i+e$], where $d_i$ and $d_i'$ refer to an original data value and the corresponding decompressed value, respectively. 
    \item Relative Error Bound (denoted by $\epsilon$). With this type of error control, the data distortion must respect the following inequality for each data point: $|d_i-d_i'|$$\leq$$\epsilon d_i$. Obviously, the smaller the original value is, the smaller the compression error it will get. The relative error bound is particularly useful to  applications requiring multiple error controls depending on the data values. 
\end{itemize}




\section{Simulation Design} \label{overview}

\begin{figure}[!t]
\centering
\captionsetup{justification=centering}
\includegraphics[width=0.4\textwidth,keepaspectratio]{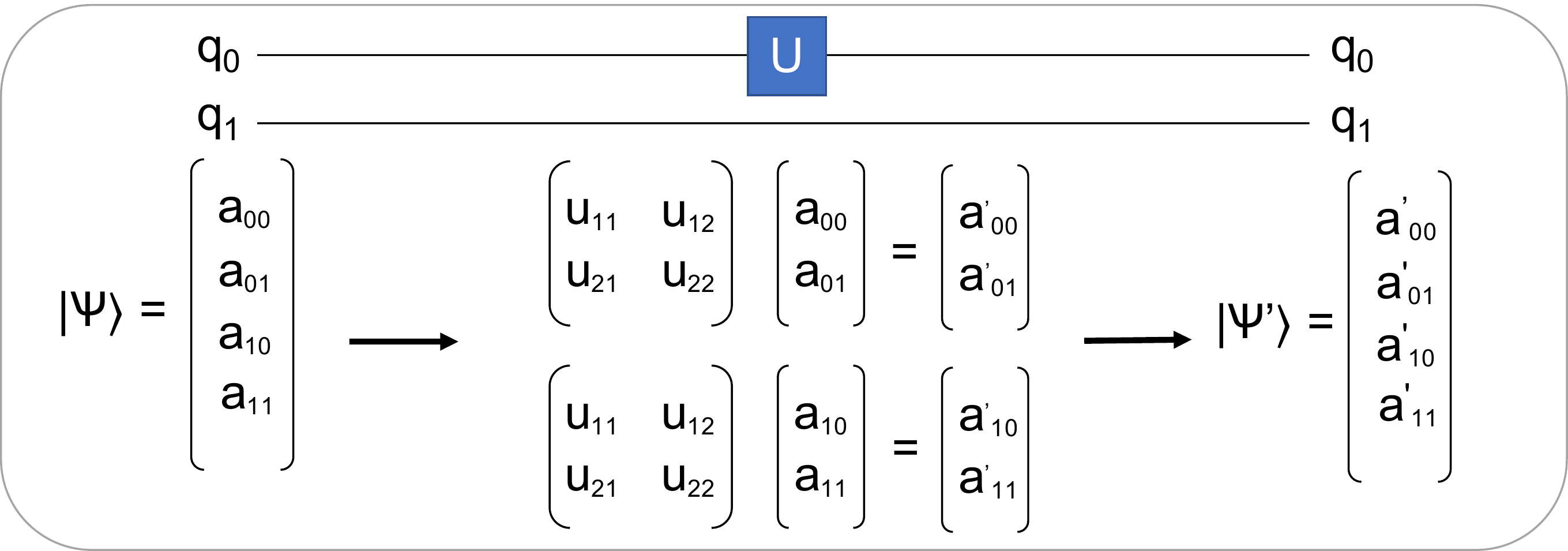}
\vspace{-3mm}
\caption{Example of two-qubit quantum state with single-qubit gate operations}
\label{fig:apply}
\end{figure}

\begin{figure*}[ht]
\centering
\captionsetup{justification=centering}
\includegraphics[width=0.7\textwidth,keepaspectratio]{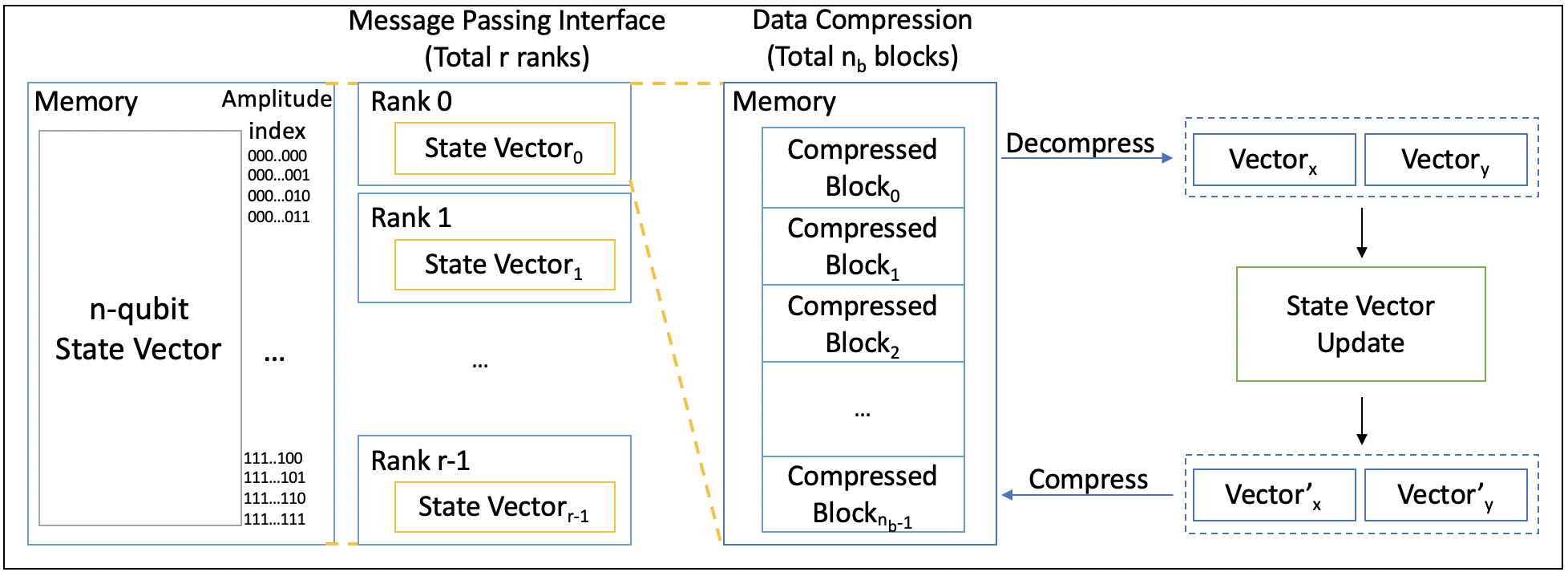}
\vspace{-3mm}
\caption{Simulation overview}
\label{fig:sim_overview}
\vspace{-3mm}
\end{figure*}

We aim to simulate general quantum circuits with high fidelity. We integrate our compression techniques into the quantum circuit simulation such that the simulation scale can be increased with the same memory capacity. Our technique allows the Schr\"{o}dinger-style simulation to trade the computation time and simulation accuracy for memory space by applying lossy compression techniques to state vectors. The lower compression error bound gives us the higher simulation fidelity, but the higher compression error bound will give us a higher compression ratio so that we can simulate the quantum circuits with larger numbers of qubits. 

\subsection{Overview of Our Simulation Flow}
To demonstrate the practicality of our design, we integrate our compression techniques into Intel-QS \cite{smelyanskiy2016qhipster}, a distributed quantum circuit simulator on a classical computer.

As mentioned in Section~\ref{background}, applying a single-qubit gate to the $k$th qubit can be represented by a unitary transformation $A  = I ^{\otimes n-k-1} \otimes U \otimes I^{\otimes k}$.  In the simulation, however, we do not need to build the entire unitary matrix $A$ to perform the gate operation. Figure~\ref{fig:apply} shows an example of applying a single-qubit gate to a two-qubit system. Applying a gate $U$ to the first qubit corresponds to applying the $2\times 2$ unitary matrix to every pair of amplitudes, whose subscript indices have 0 and 1 in the first bit and all remaining bits are the same. In the same way, performing a single-qubit gate to the second qubit is to apply the unitary to every pair of amplitudes whose subscript indices differ in the second bit. More extensively, applying a single-qubit gate to the $k\text{-th}$ qubit of an $n$-qubit quantum system is to apply the unitary to every pair of amplitudes whose subscript indices have 0 and 1 in the $k\text{-th}$ bit, while all other bits remain the same.
\begin{equation}
\label{eq:matrix1}
\begin{bmatrix}a'_{*...*0_k*...*}\\a'_{*...*1_k*...*}\end{bmatrix} = \begin{bmatrix}u_{11} & u_{12}\\u_{21} & u_{22}\end{bmatrix}\begin{bmatrix}a_{*...*0_k*...*}\\a_{*...*1_k*...*}\end{bmatrix}
\end{equation}

As for a generalized two-qubit controlled gate, the unitary is applied to a target qubit $t$ if the control qubit $c$ is set to $\ket{1}$; otherwise $t\text{th}$ is unmodified.
\begin{equation}
\label{eq:matrix2}
\begin{bmatrix}a'_{*1_c...*0_t*...*}\\a'_{*1_c...*1_t*...*}\end{bmatrix} = \begin{bmatrix}u_{11} & u_{12}\\u_{21} & u_{22}\end{bmatrix}\begin{bmatrix}a_{*1_c...*0_t*...*}\\a_{*1_c...*1_t*...*}\end{bmatrix}
\end{equation}

Figure~\ref{fig:sim_overview} shows an overview of our simulation design. The Message Passing Interface (MPI) \cite{gropp1999using} is used to execute the simulation in parallel.

Assuming we simulate $n$-qubit systems and have $r$ ranks in total, the state vector is divided equally on $r$ ranks. To reduce the memory requirement, we further divide the partial state vector into $n_b$ blocks on each rank. Each block is stored in  compressed format on the memory. To complete a gate operation, we need to apply matrix multiplication to the pair of amplitudes whose subindices are 0 and 1 at the target qubit position, so at most two blocks are decompressed, $Vector_x$ and $Vector_y$, and then update the state vectors. After all the amplitudes in $Vector_x$ and $Vector_y$ are updated, we compress the state vectors and move to the next two blocks. Once all the blocks have been updated, a gate operation is completed. 

In this way, the total number of bytes required for the simulation of a rank is as follows: 
\begin{equation}
\label{eq:totalnbBytes}
nbBytes = \sum_i sizeof(CB_i) + 2(\frac{2^{n+4}}{r \times n_b}),
\end{equation}

where $CB_i$ is the $i$th compressed block. 

\subsection{MCDRAM Memory Configuration}
Multichannel DRAM (MCDRAM) is a high-bandwidth, low-capacity memory, packaged with the Intel Xeon Phi processor \cite{sodani2015knights}.
Because  several repeated compression and decompression operations are involved in the simulation, we can achieve the performance improvement by utilizing  MCDRAM. Every time  the block is decompressed, we  decompress the state vectors to MCDRAM. To allocate memory in MCDRAM, we use \emph{mkl\_malloc} function to acquire memory space. This memory allocation strategy directly improves the performance of compression, gate operation, and decompression. To achieve this, we set the machine to \emph{equal} mode, 50\% cache and 50\% flat.

\subsection{Integration Details}
We build our compressor as a C library and add it into the Intel-QS building process.
In the initialization of the simulation process, we create the state vectors in blocks, and compress them as compressed blocks. During the simulation, if the process needs to update the state vector, it calls our compressor library to decompress the block to the pre-allocated MCDRAM. After the state vector update, the block is compressed, and the process moves to the next block.

\begin{figure}[ht]
\centering
\captionsetup{justification=centering}
\includegraphics[width=0.44\textwidth,keepaspectratio]{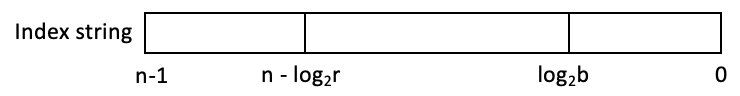}
\vspace{-3mm}
\caption{Amplitude index segments}
\label{fig:index_segments}
\end{figure}

Since we allow decompression of only two blocks for each rank at the same time, we must select the corresponding blocks to be decompressed. For $n$-qubit systems, assuming we have $r$ ranks and each block contains $b$ amplitudes, the amplitude index string can be divided into three segments (Figure~\ref{fig:index_segments}). When we execute a single-qubit gate computation on the target qubit position $q$, we find the corresponding blocks according to the segment $q$ belongs to. 

\begin{itemize}
\item $q < log_2b:$ Both amplitudes are in the same block.
\item $n-log_2r \leq q \leq log_2b:$ Both amplitudes are in the same rank but with different blocks. We can use the number $q$ to find the corresponding blocks in the rank.
\item $q > n-log_2r:$ The pair of amplitudes are in the different ranks. The blocks have to be exchanged between different ranks.
\end{itemize}

Two-qubit gate operations are similar to single-qubit gate operations, but we  modify the amplitudes only when the control qubit is set to $\ket{1}$. According to the position of the control qubit $c$, we also have three cases to determine whether the amplitudes should be modified:
\begin{itemize}
\item $c < log_2b:$ If $c$th bit is 0, the amplitude is left unmodified.
\item $n-log_2r \leq c \leq log_2b:$ If $c$th bit is 0, the whole block is left unmodified.
\item $c > n-log_2r:$ If $c$th bit is 0, the whole rank is left unmodified.
\end{itemize}

\subsection{Compressed Block Cache}
For most of the quantum circuits, the amplitudes may share the same value \cite{zulehner2018advanced}. By exploiting redundancies in the quantum state, we can reduce the computation time significantly by constructing a compressed block cache. 

\begin{figure}[ht]
\centering
\captionsetup{justification=centering}
\includegraphics[width=0.36\textwidth,keepaspectratio]{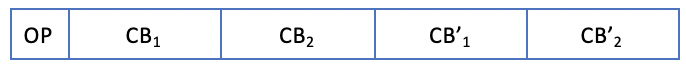}
\vspace{-3mm}
\caption{Compressed block cache line}
\label{fig:cbc}
\end{figure}

Figure~\ref{fig:cbc} shows the contents of a compressed block cache line. The first element $(OP)$ is the gate operation and the target qubit position. The second and third elements $(CB_1, CB_2)$ are the compressed blocks before the gate operation. The rest of the elements $(CB'_1,CB'_2)$ are the compressed blocks after the gate operation.
Each rank maintains a memory space for the compressed block cache with 64 cache lines. When a gate operation is executed, our computation procedure first checks whether the pattern $(OP, CB_1, CB_2)$ is in the cache. If there is a cache hit, then the computation is done by directly returning the blocks  $(CB'_1,CB'_2)$, and thus the performance is improved by reducing the compression, computation, and decompression time.

The cache replacement policy is least recently used. If there is no redundancy in the quantum state, the cache hit rate will be 0, and this will introduce the cache miss penalty in our simulation. Thus, our simulator will disable the compressed block cache if the cache hit rate is always zero.

\subsection{Simulation Checkpoint} \label{sec:checkpoint}
In  general,  most    supercomputing  systems  have  a  24-hour  wall-time  limit.  This  runtime  constraint  puts  a  circuit  depth  limit  on the simulation. However, one can save the compressed blocks before terminating the job  and  then  resume  the  task  by  loading  the  compressed blocks in the next job submission.

\subsection{MPI Configuration}
\begin{figure}
\centering
\includegraphics[scale=.43]{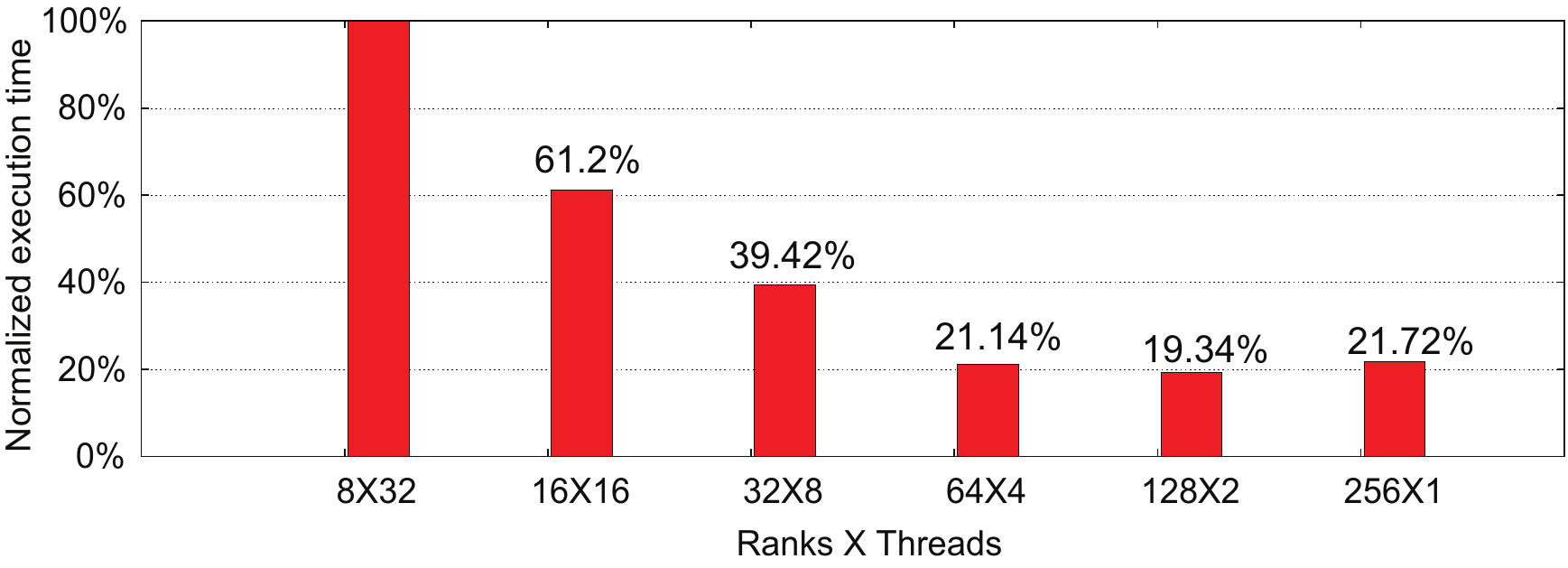}
\vspace{-3mm}
\caption{Normalized execution time for running 35-qubit random circuit simulations.}
\vspace{-2mm}
\label{fig:ranks}
\end{figure}

To understand the performance under different MPI rank configuration, we run the 35-qubit random circuit simulation with various ranks per node (Figure~\ref{fig:ranks}). On the KNL nodes, each node has 64 cores and 256 threads. We found that the setting of 128 ranks per node gives the best performance.

\subsection{Variable Error Bound Compression}
\label{sec:err-bound-cmpr}
To keep the simulation accuracy, we use the lossless compression Zstd\cite{zstd} in the beginning of the simulation when the compression ratio is still high enough to fit all compressed blocks in the memory space. During the simulation, the quantum state becomes more and more complicated, and hence the lossless compression ratio is lower. When the compression ratio is too low to fit all compressed blocks in the memory, our simulation will use lossy compression to compress the state vectors. To control the error, we use a pointwise relative error bound to compress the state vector. This compression mode guarantees that the decompressed data point $|D'|$ must be in the range of $(|D(1-\delta)|, |D|)$, where $D$ is the original data and $\delta$ is the error bound. In this work, we have five different levels of  error bounds: 1E-5, 1E-4, 1E-3, 1E-2, and 1E-1. Whenever the compression ratio is not enough, the error bound is relaxed to the next level (larger error). We  give a detailed discussion about our lossy compression technique in the next section.

\subsection{Lower Bounds on Simulation Accuracy}
Since lossy compression is used in our simulations, information is lost with every lossy compression, causing a decrease in the simulation's overall accuracy. The accuracy of our simulation can be quantified by the state fidelity, a measure of the similarity of two quantum states \cite{nielsen2002quantum}.

The fidelity takes values between $0$ and $1$, with higher fidelity indicating greater similarity between the two states. A fidelity of $1$ would indicate that the two quantum states are the same. The pure state fidelity between the ideal output state, $\psi_{ideal}$, and the output state from our simulation, $\psi_{sim}$, can be described by the following simplified equation \cite{nielsen2002quantum}.
\begin{equation}
\label{eq:idealsim}
F(ideal, sim) = |\langle \psi_{ideal}| \psi_{sim} \rangle| 
\end{equation}

\begin{figure}
\centering
\vspace{-6mm}
\includegraphics[scale=.58]{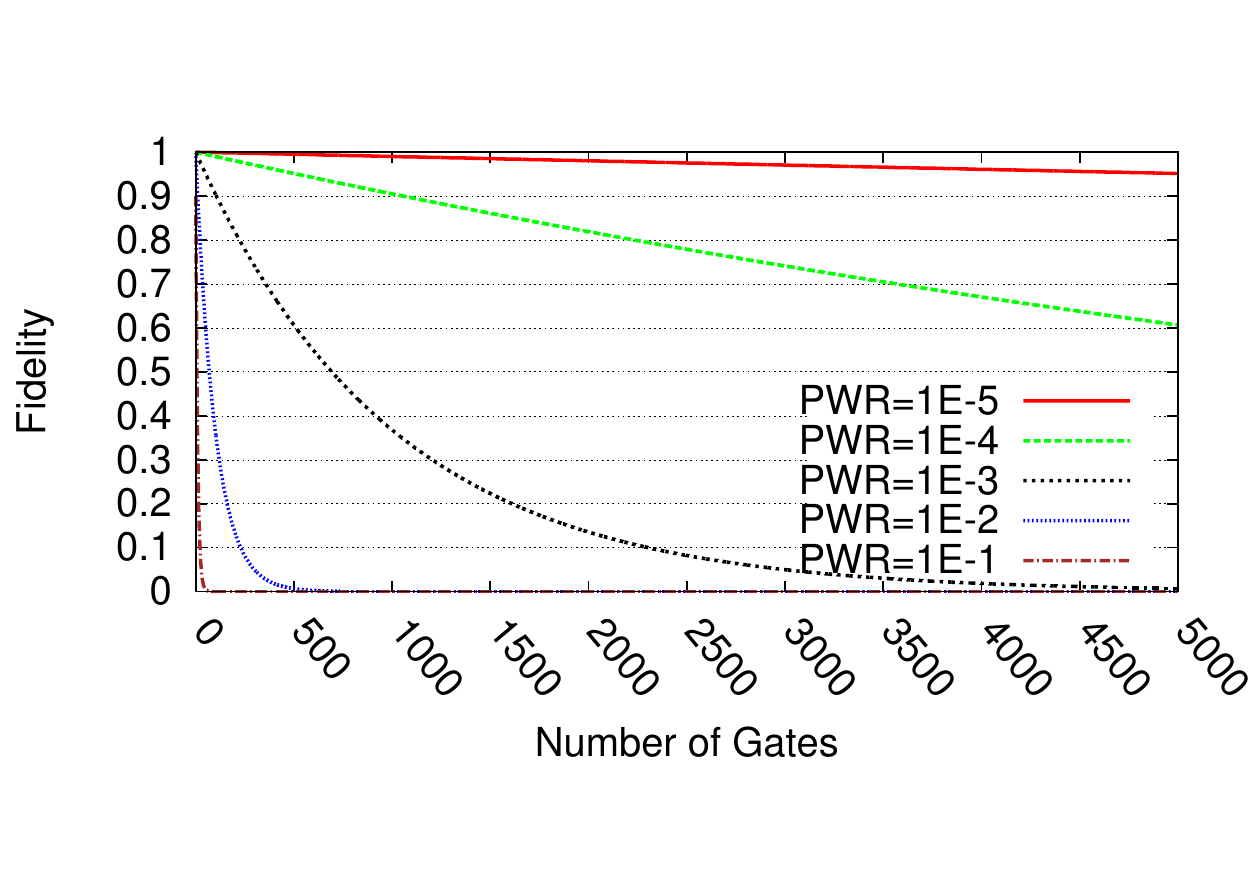}
\vspace{-9mm}
\caption{Minimum bounds for fidelity with an increasing number of gates at different error levels}
\vspace{-3mm}
\label{fig:fidelity}
\end{figure}

Since the errors are bounded in our simulation, the fidelity can be estimated by propagating the maximum error bounds at each gate to calculate the maximum decrease in fidelity (from the ideal fidelity of 1) that comes from each lossy compression and then finding their combined impact on the overall maximum decrease in fidelity. 
Suppose that a gate has a percentage error of $\delta$. Then if a complex amplitude $a_i$ in the ideal state vector is represented as $a+bi$, the corresponding amplitude after our lossy compression in the simulation can be represented by $a'+b'i$, where $|a'| \geq |a(1-\delta)|$ and $|b'| \geq |b(1-\delta)|$. 
Thus, we see that the state fidelity can be calculated as follows:
\vspace{-2mm}
\begin{equation}
F(ideal, sim) = |\langle \psi_{ideal}| \psi_{sim} \rangle| \geq \sum_{i=0}^{2^n-1} a_i^2 (1-\delta) = 1-\delta.
\vspace{-2mm}
\end{equation}
\\So we see that the minimum fidelity drops by a factor of $ (1-\delta)$ after applying a lossy compression with a maximum percentage error bound of $\delta$. Suppose that for the $i$th gate, the error bound is $\delta_{i}$ on the lossy compression. Then the lower bound on the fidelity after applying this compression will drop to $(1-\delta_{i})$ times the lower bound on the fidelity calculated before lossy compression on this gate. 
Combining the contributions of all the gates in the simulation allows us to calculate the lower bound on the simulation fidelity as
\begin{equation} 
 F(ideal, sim) \geq \prod_{i}{(1-\delta_{i})}.
\end{equation}
 If all the  $\delta_{i}$ were 0, the simulation fidelity would be 1, which makes sense because our simulation would then be identical to the simulation without lossy compression against which we are calculating the fidelity.
As mentioned before, in our simulation with lossy compression, the error bounds $\delta_{i}$ can be set to 0 (lossless), 1E-5, 1E-4, 1E-3, 1E-2, and 1E-1. 
Figure~\ref{fig:fidelity} shows how fidelities change with the number of gates when different error levels are applied.

\section{Adaptive Compression Optimized for Quantum Circuit Simulation}
\label{sec:compression}

In this section, we propose a novel error-bounded compression method that can significantly control memory footprint for full-state quantum circuit simulations that were unreachable before. 

\subsection{Assessment of Existing Error-Bounded Lossy Compressors}
\label{sec:best-model}

First, we explore the best solution from among the existing compressors for the quantum circuit simulations.  
We choose SZ \cite{sz,sz21-msst19}, FPZIP \cite{fpzip}, and ZFP \cite{zfp} in our exploration because they have been confirmed as the best error-bounded compressors on many scientific datasets \cite{sz,sz21-msst19,model-ipdps18}.

We perform the assessment using two well-known quantum circuits: a quantum approximate optimization algorithm (QAOA) \cite{farhi2014quantum} and the random circuit proposed by Google to show quantum supremacy \cite{boixo2018characterizing}. In this analysis, we run 36 qubits for both circuits and denote them as qaoa\_36 and sup\_36, respectively. 

As discussed in Section \ref{sec:cmpr-tech-background}, there are two types of error bounds, both of which have been widely used by scientific applications, so we evaluate the compression ratios for quantum circuit simulation data based on both types of errors. Since each rank involves hundreds or thousands of data blocks each having different value ranges, we perform the compression based on the absolute error bound in the regard of value range in our evaluation, without loss of generality. That is, we set the absolute error bound to a fixed percentage of the value range in each block. For instance, 1E-2 means 1\% of the value range in the following figures.

Figure \ref{fig:sz-zfp-results-abs} presents the compression ratios of SZ and ZFP based on different absolute error bounds. FPZIP is missing in this figure because it does not support an absolute error bound. We can see  that SZ always leads to one or two orders of magnitude higher compression ratios than ZFP does at every error bound. For instance, on qaoa\_36, SZ can lead the compression ratios up to about 100:1, while ZFP's compression ratios are always less than 10:1. For the dataset sup\_36, the compression ratios of SZ and ZFP are about 28$\sim$126 and 4.25$\sim$12.6, respectively.

\begin{figure}[ht] \centering

\hspace{-7mm}
\subfigure[{qaoa\_36}]
{
\includegraphics[scale=.38]{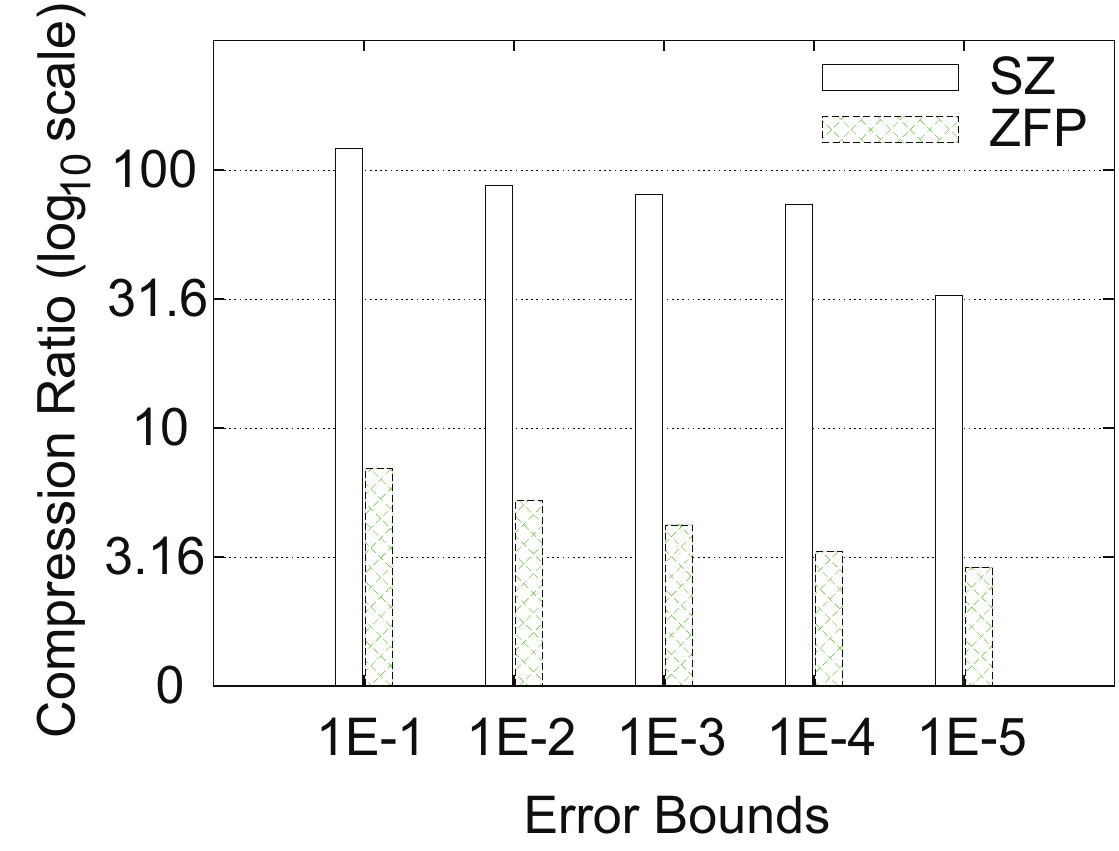}
}
\hspace{-3mm}
\subfigure[{sup\_36}]
{
\includegraphics[scale=.38]{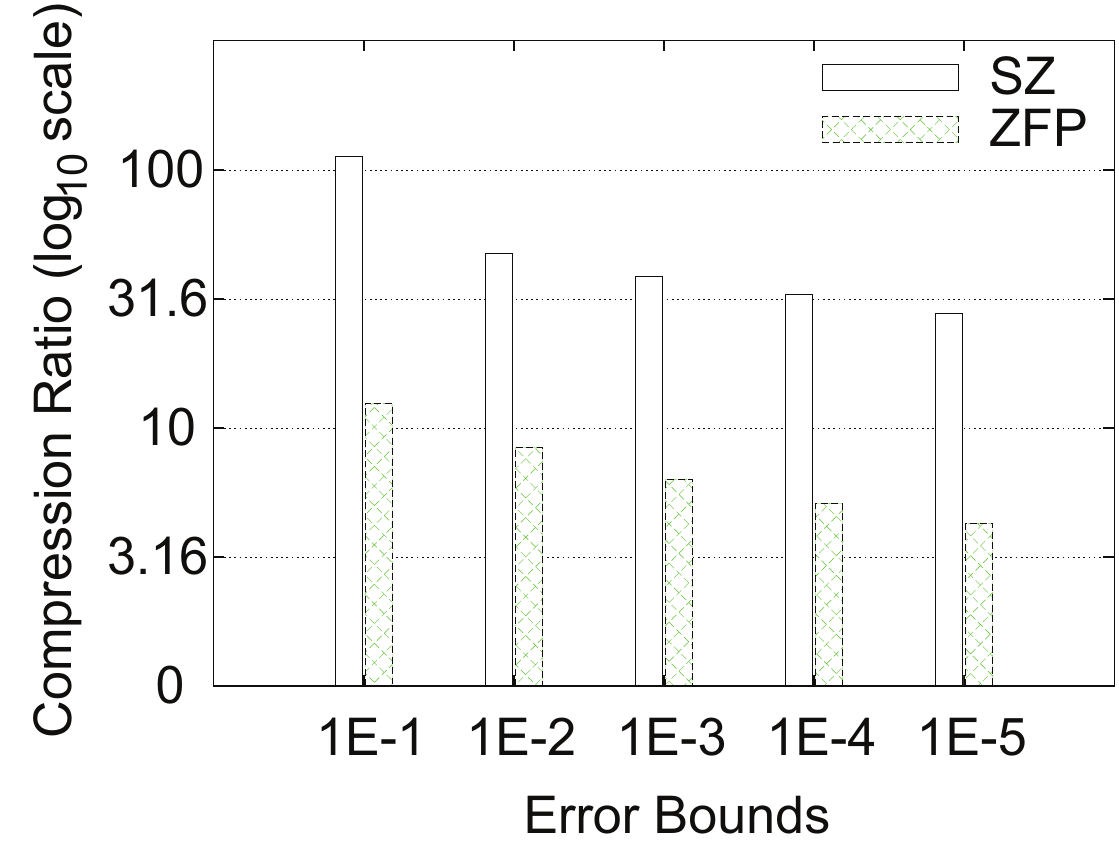}
}
\hspace{-8mm}

\vspace{-4mm}
\caption{Compression ratio of SZ vs. ZFP (absolute error)}
\label{fig:sz-zfp-results-abs}
\end{figure}

\begin{figure}[ht] \centering

\hspace{-7mm}
\subfigure[{qaoa\_36}]
{
\includegraphics[scale=.38]{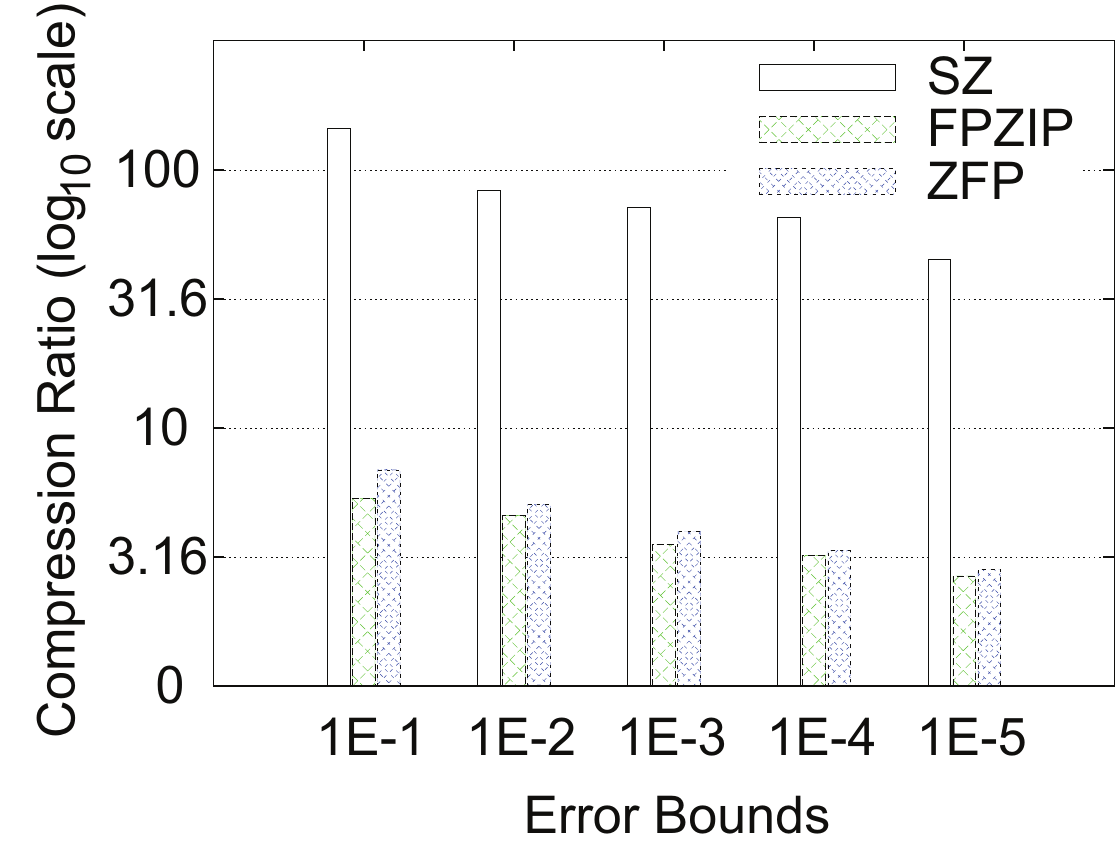}
}
\hspace{-3mm}
\subfigure[{sup\_36}]
{
\includegraphics[scale=.38]{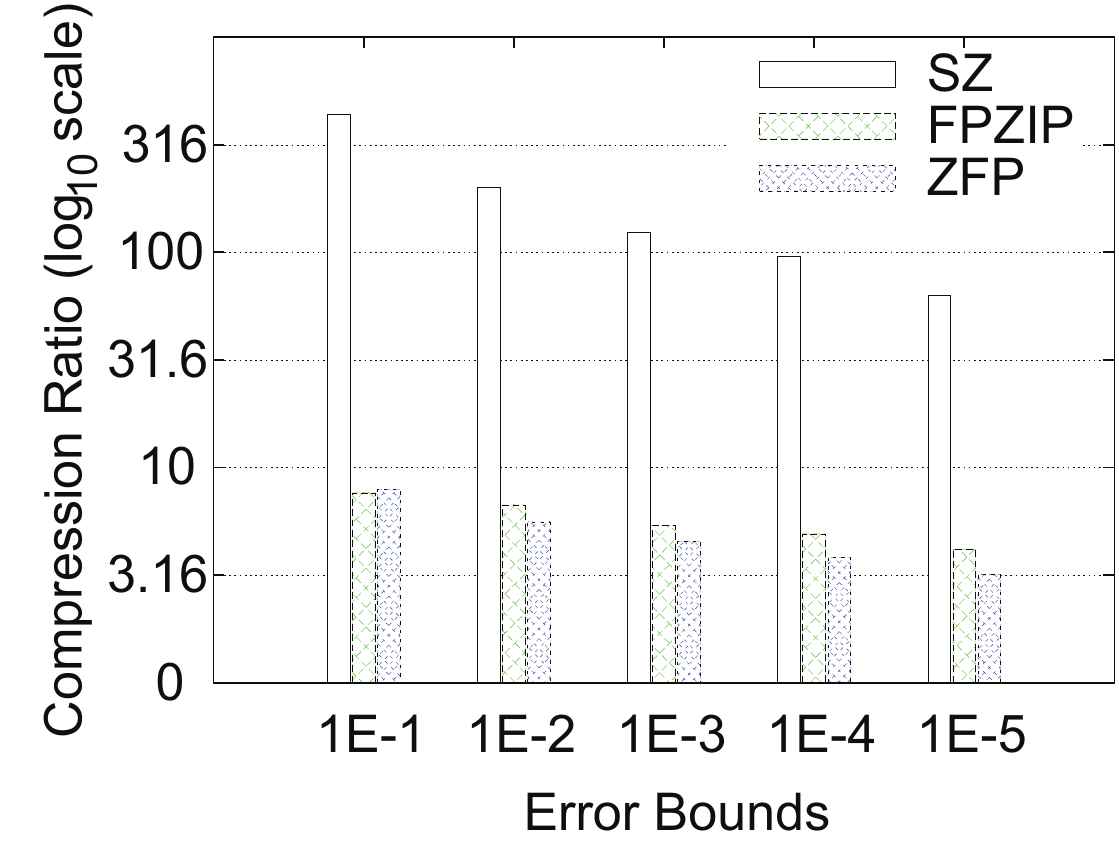}
}
\hspace{-8mm}

\vspace{-4mm}
\caption{Compression ratio of SZ,FPZIP,ZFP (relative error)}
\label{fig:sz-zfp-results-rel}
\end{figure}

For the pointwise relative error bound, for ZFP, we transform the original data to the logarithm domain and then compress the transformed data with absolute error bounds, for fairness of the comparison. Such a log-preprocessing-based compression has been validated as the best way to do the pointwise relative-error-bounded compression \cite{xin2018}.
As for FPZIP, it provides a so-called precision number (=4$\sim$64) to control the pointwise relative error bound. The larger the precision number is, the lower the pointwise relative error bound obtained. We set the precisions to 16, 18, 22, 24, and 28 for FPZIP in our experiments because they correspond to the pointwise relative error bounds of 1E-1, 1E-2, 1E-3, 1E-4, and 1E-5 approximately.  
Figure \ref{fig:sz-zfp-results-rel} clearly shows that SZ always leads to much higher compression ratios than do the other two compressors with the same pointwise relative error bounds. 

\begin{figure}[ht] \centering

\hspace{-10mm}
\subfigure[{qaoa\_36}]
{
\includegraphics[scale=.4]{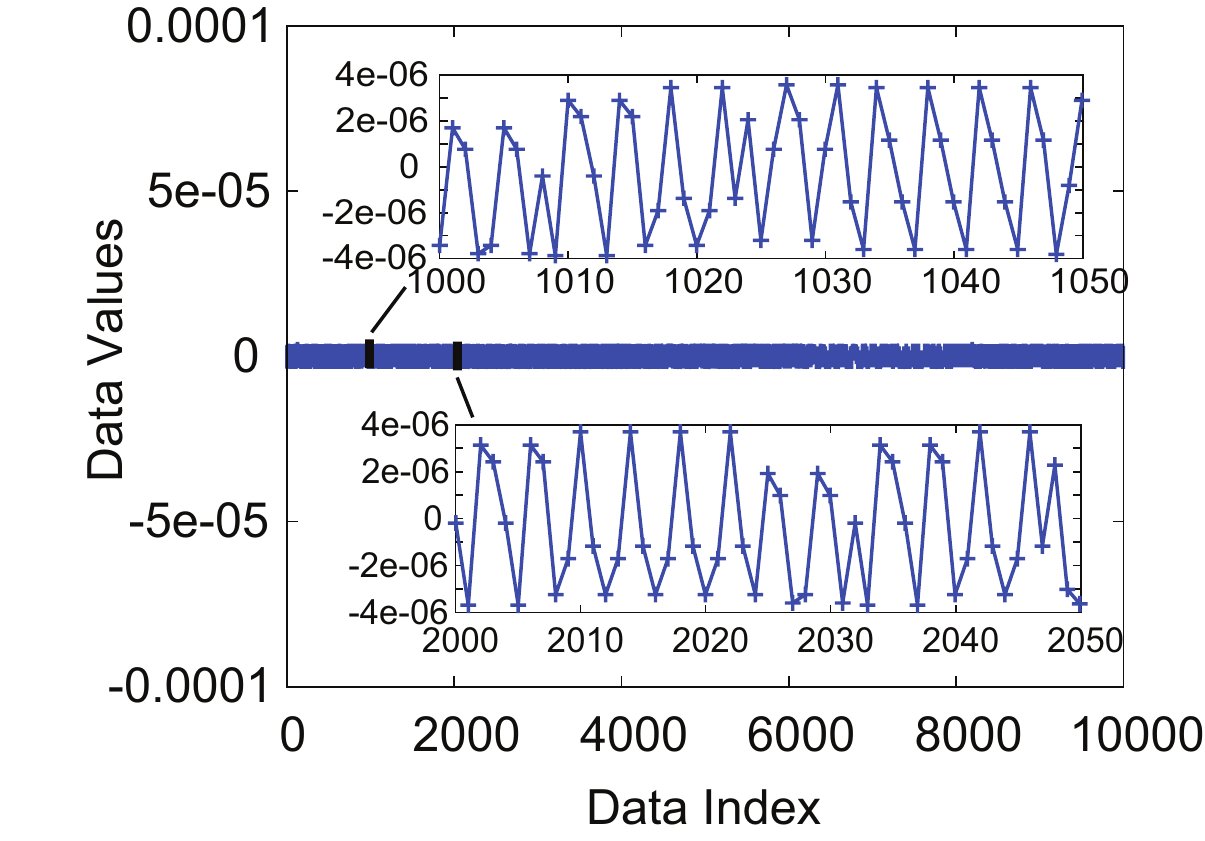}
}
\hspace{-9mm}
\subfigure[{sup\_36}]
{
\includegraphics[scale=.4]{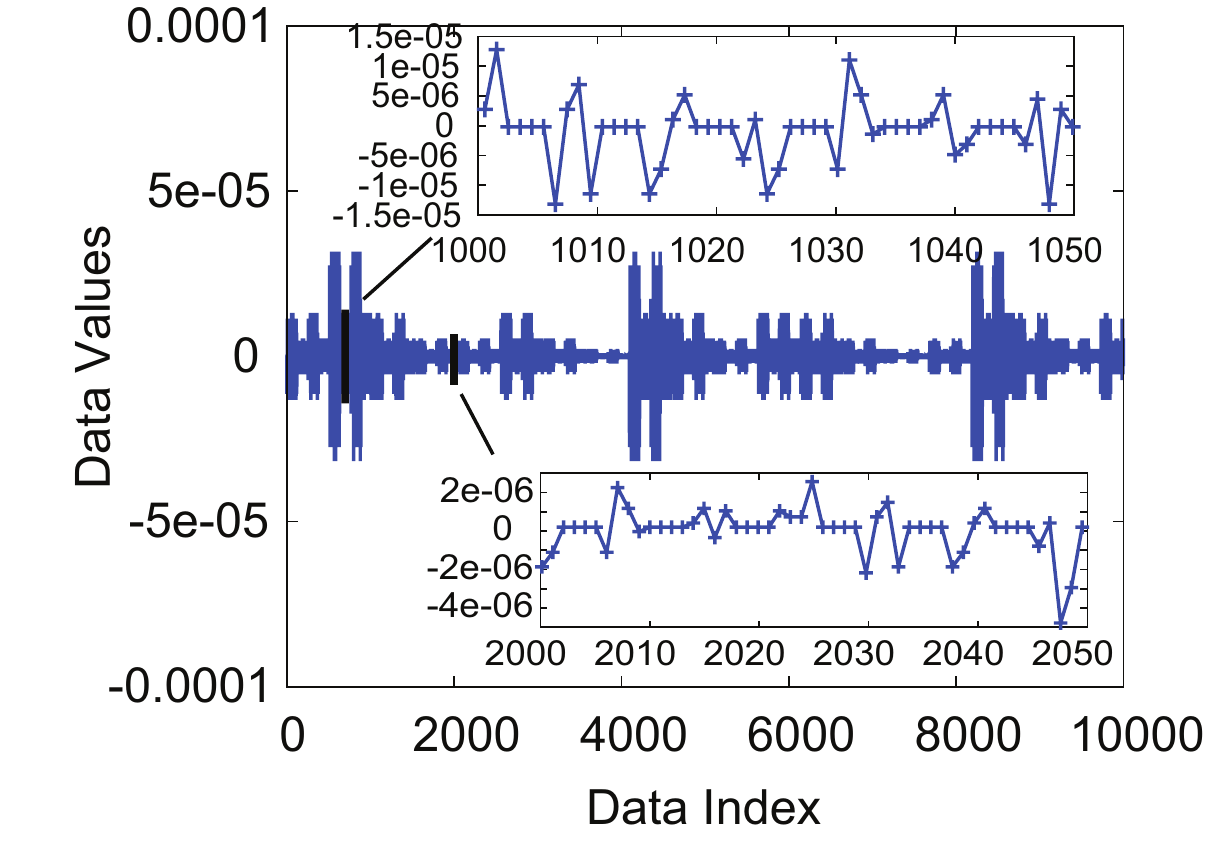}
}
\hspace{-10mm}

\vspace{-4mm}
\caption{Illustration of Value changes of quantum circuit simulation data: the data exhibit a high spikiness such that existing lossy compressors cannot work effectively.}
\label{fig:illus-data}
\end{figure}

Based on our analysis, we conclude that SZ is superior to the other two compressors. The key reason is  as follows. For ZFP, it substantially replies on the high smoothness of data, especially when the error bounds are set to a relatively low value. However, the quantum simulation data are not smooth at all, as illustrated in Figure \ref{fig:illus-data}, such that the domain-transform in ZFP would totally lose its effectiveness, leading to poor compression ratios. The key difference between FPZIP and SZ is that the former adopts a totally different encoding method, unlike SZ adopting linear-scaling quantization + Huffman encoding + Zstd. In what follows, we treat SZ as the baseline and propose a new lossy compression method that is more effective on the quantum circuit simulation data.

\subsection{Optimizing the Compression Strategy}
\label{sec:cmpr-opt}


In our work, we adopt a hybrid, adaptive compression pipeline to control the memory footprint while keeping the high fidelity of the simulation results and relatively low time overhead. Specifically, we observe that at the early stage of the simulation, a large majority of the data are zero. In this situation, the lossless compressor Zstd can also get a satisfactory compression ratio, without any loss of data fidelity. Although SZ leads to much higher compression ratios in this situation, it may introduce data distortion,  causing degraded fidelity unexpectedly. As the simulation goes on, however, the simulation data will become more and more complicated, such that Zstd will suffer from low compression ratios. In this situation, the error-bounded lossy compressor is  helpful in controlling the memory footprint while keeping a high fidelity of simulation results. Thus, we adopt both Zstd and error-bounded lossy compression during each entire simulation. More details are in Section \ref{sec:err-bound-cmpr}.

Developing a fast, effective error-bounded lossy compressor is critical to the overall simulation results (including fidelity and simulation time). In particular, we need to use a pointwise relative error bound to do the compression, as discussed in Section \ref{overview}. To this end, we explored a battery of novel compression strategies to optimize both the compression ratios and compression speed for quantum circuit simulations, as described below.\\ 

\noindent\textbf{Solution A (SZ 2.1)}: This solution is the state-of-the-art error-bounded lossy compressor - SZ \cite{sz,tao2017significantly,sz21-msst19}. In the context of quantum circuit simulations, the data can  be treated only as a 1D array. In SZ 2.0, the pointwise relative-error-bounded compression involves the following steps: (1) logarithm data transform; (2) absolute-error-bounded compression on log-transformed data (Lorenzo prediction \cite{lorenzo} + quantization \cite{tao2017significantly}); (3) Huffman encoding; and (4) Zstd lossless compression. Since log-transform is expensive, the SZ development team developed SZ 2.1 leveraging a table lookup method to accelerate the compression significantly \cite{sz21-msst19}, without degrading the compression ratios. However, the compression/decompression speed still cannot meet the expected level for quantum circuit simulations (to be shown later), which motivates us to further explore a new, faster compression method instead.\\

\noindent\textbf{Solution B (SZ 2.1 with complex type supported)}: Quantum state amplitudes are stored in the form of complex data type, in which the real numbers and imaginary numbers are stored alternatively, such that the prediction accuracy would be degraded to a certain extent, leading to limited compression ratios. Accordingly, Solution B  improves the prediction accuracy by performing the prediction on the real numbers and imaginary numbers, respectively. In addition, we set the maximum number of quantization bins to 16,384 (unlike the default setting of 65,536 in SZ 2.1), which can significantly improve the compression/decompression rate (to be shown later).\\

\noindent\textbf{Solution C (XOR leading-zero data reduction + bit-plane truncation + Zstd)}: Solution B can improve the compression ratios in some cases, but its compression speed is still lower than expected such that the total simulation would slow  significantly compared with the original compression-free execution. To reduce the compression/decompression time significantly, we developed Solution C, a simple yet efficient compression pipeline that is particularly suitable for quantum circuit simulation. This method involves three key steps. For each data point, it first leverages \emph{XOR leading-zero data reduction method} \cite{fpc,sz}, which uses a two-bit code to record the number of exactly the same bytes between the current value and its preceding value. 
    Then the algorithm truncates the insignificant bit-planes based on the required relative error bound $\epsilon$. Specifically, the significant number of bits can be calculated as follows. 
    \begin{equation}
    \label{eq:sigbits}
    Sig\_Bit\_Count = Bit\_Count(Sign\&Exp) - EXP(\epsilon),
    \end{equation}
    where $EXP(\epsilon)$ refers to the exponent of the relative error bound $\epsilon$ (e.g., $EXP(0.01)$=$-7$) and $Bit\_count(Sign\&Exp)$ is  the total number of bits used to represent sign and exponent in the IEEE 754 format (e.g., it is equal to 12 for double precision). We then adopt Zstd lossless compression to shrink the data significantly.\\

\noindent\textbf{Solution D (Reshuffle + Solution C)}: Since the simulation data are stored in the complex data type (i.e., real number and imaginary number alternatively), one plausible idea is to  reorganize the data into real numbers and imaginary numbers separately before compressing the data. Such a reshuffle step may help improve compression ratio especially when the real numbers and imaginary numbers are located in different non-overlapped value ranges. The reason is that in Solution C, the last step Zstd involves a dictionary-matching stage (LZ77 \cite{lz77}) leveraging potential repeated patterns in the data streams and the reshuffling step that separates the real numbers and imaginary numbers may improve the pattern matching to a certain extent. Accordingly, we developed Solution D, which might have higher compression ratios with a slightly lower compression/decompression rate.

\begin{figure}[ht] \centering

\hspace{-7mm}
\subfigure[{qaoa\_36}]
{
\includegraphics[scale=.36]{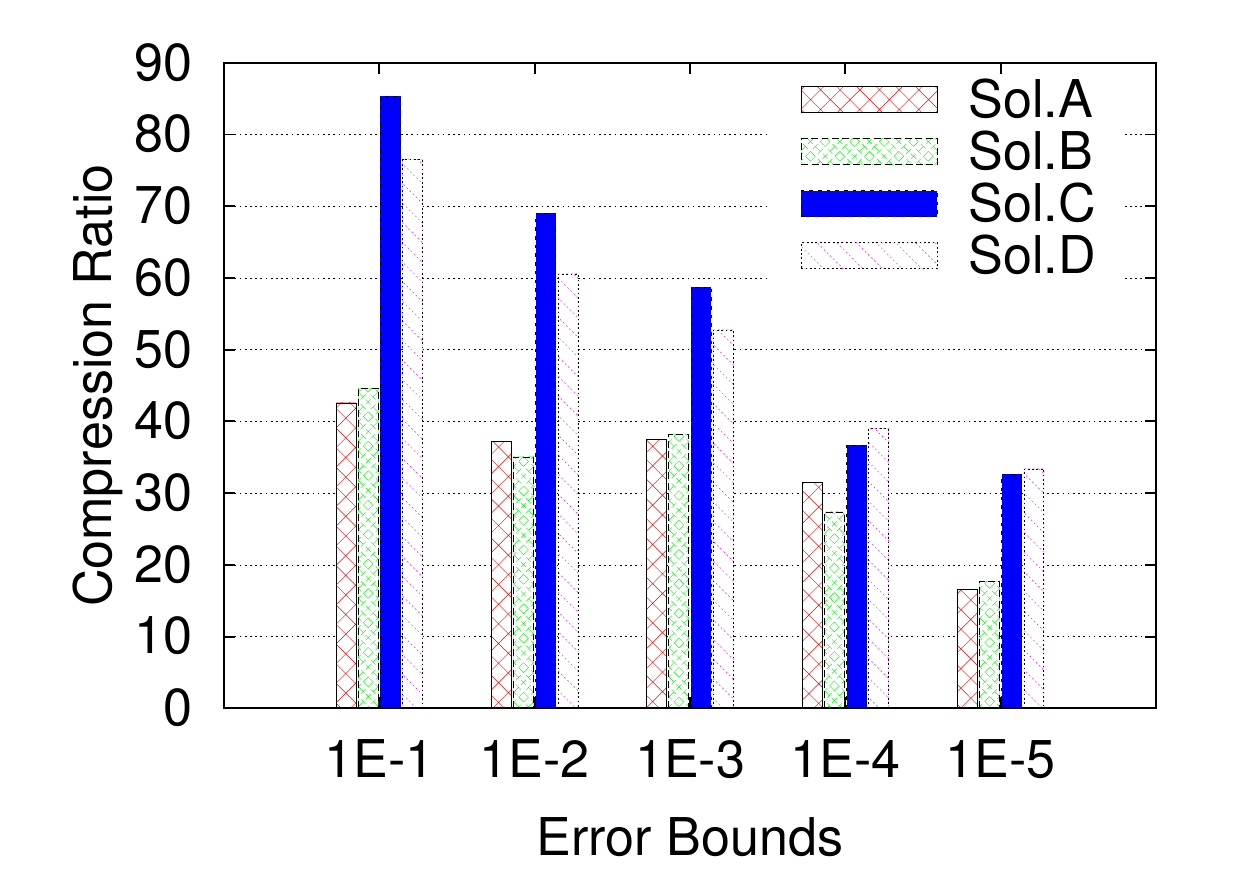}
}
\hspace{-7mm}
\subfigure[{sup\_36}]
{
\includegraphics[scale=.36]{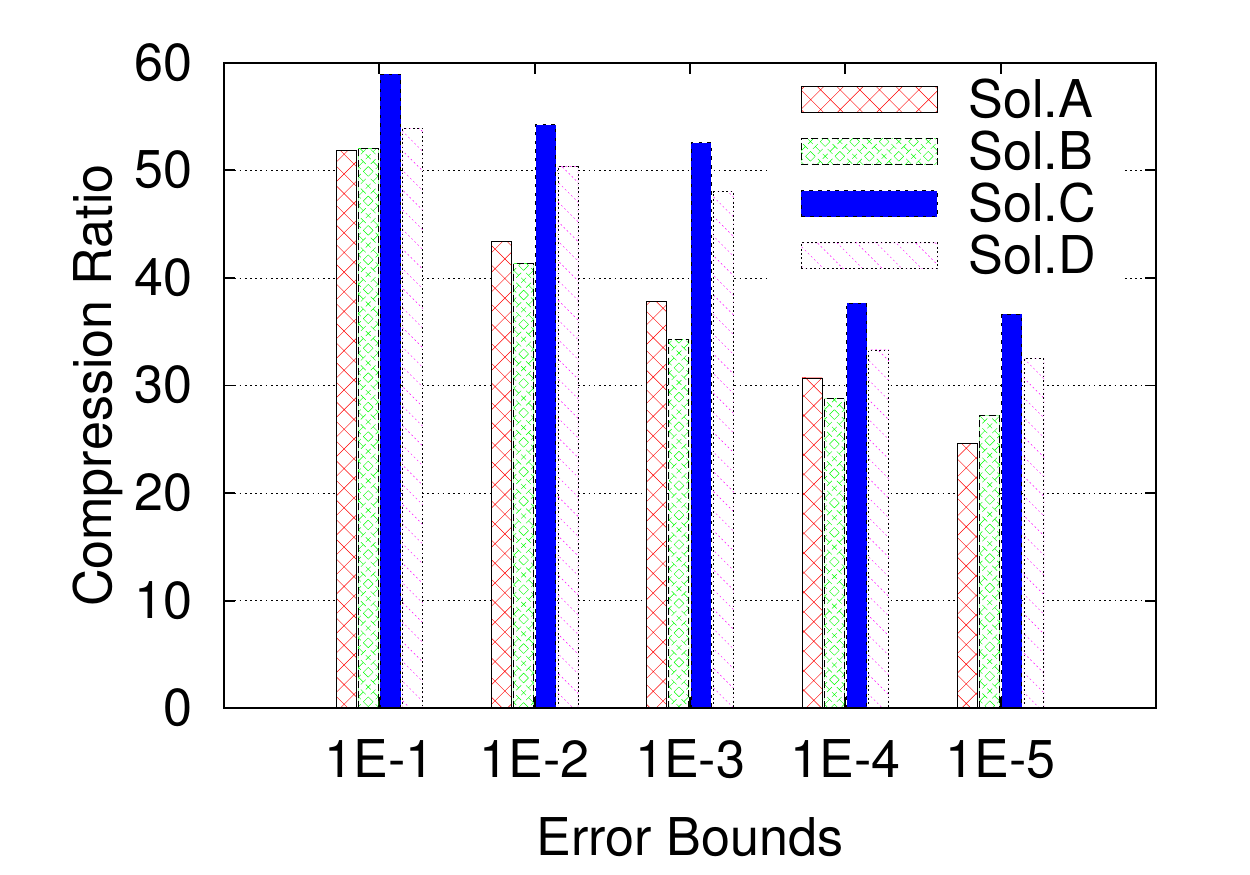}
}
\hspace{-8mm}

\vspace{-4mm}
\caption{Compression ratio of 4 solutions (relative error)}
\label{fig:4sol-cr}
\end{figure}

We evaluate all the four solutions by doing the compression and decompression with the two simulation datasets qaoa\_36 and sup\_36. Figure \ref{fig:4sol-cr} presents the compression ratios of the four solutions. We can see that the classic compressor SZ 2.1, either supporting complex type (Solution B) or not (Solution A), suffers from about 30\%$\sim$50\% lower compression ratios than do Solutions C and D. The likely reason is that the simulation data exhibit spiky changes (as illustrated in Figure \ref{fig:illus-data}) such that the SZ compressor always suffers from low prediction accuracy. Besides, we note that the solution C may lead to slightly higher compression ratios than does the solution D in some cases, which also makes sense as explained as follows. Note that the only difference between these two solutions is the extra reshuffle step in the solution D. This step might affect the compression ratio only because of the LZ77 stage in the last step Zstd of the two solutions, as analyzed previously. With this in mind, the compression ratio actually may not change significantly in between since the pattern matching efficiency for the two solutions could be very similar. On the one hand, Zstd improved LZ77 by adopting a pretty large window size. On the other, the real numbers and imaginary numbers are of the similar value ranges (as shown in Figure \ref{fig:illus-data}), such that the reshuffling step may not induce more regular data. In this sense, the solution C and D can be deemed having comparative compression ratios on the QC simulation datasets.   

\begin{figure}[ht] \centering
\vspace{-2mm}
\hspace{-7mm}
\subfigure[{qaoa\_36 (Cmpr)}]
{
\includegraphics[scale=.36]{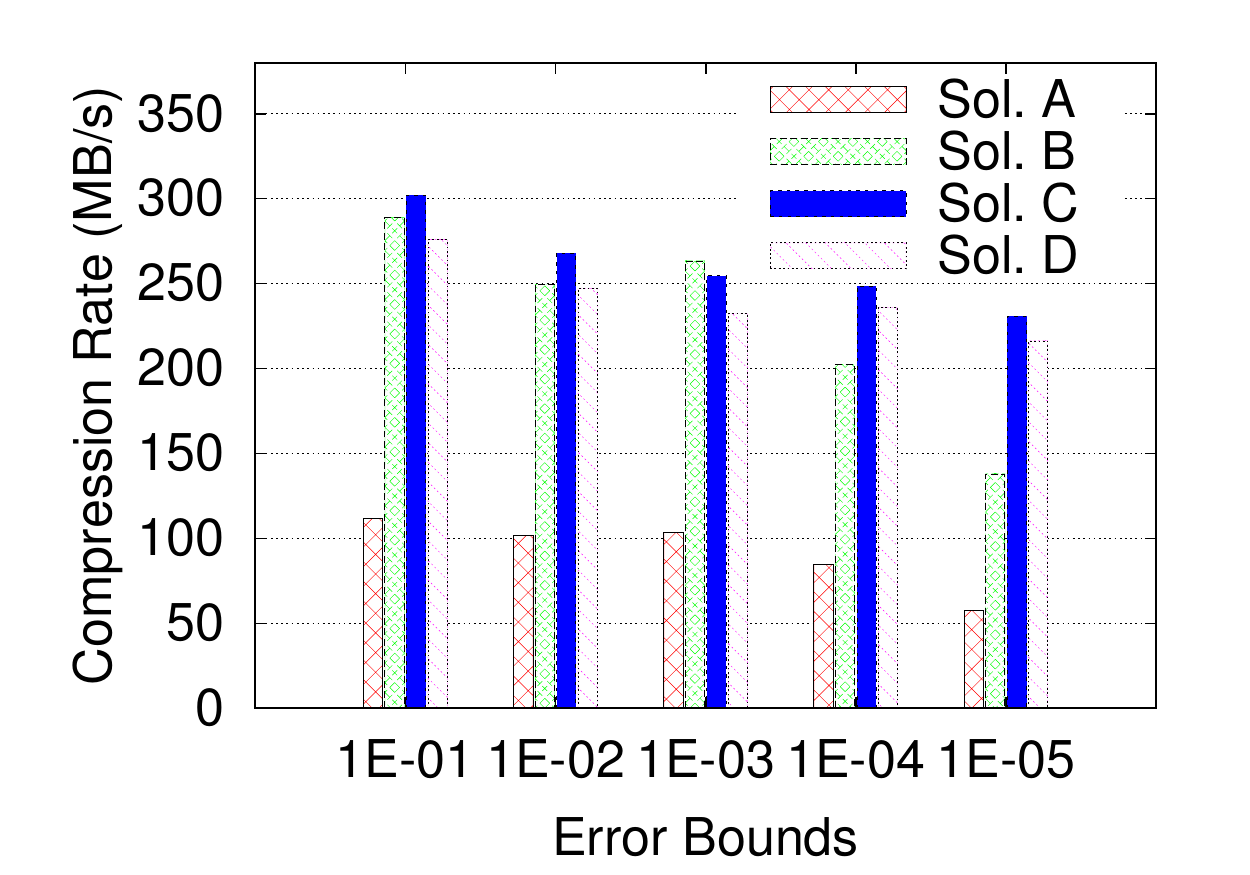}
}
\hspace{-7mm}
\subfigure[{sup\_36 (Cmpr)}]
{
\includegraphics[scale=.36]{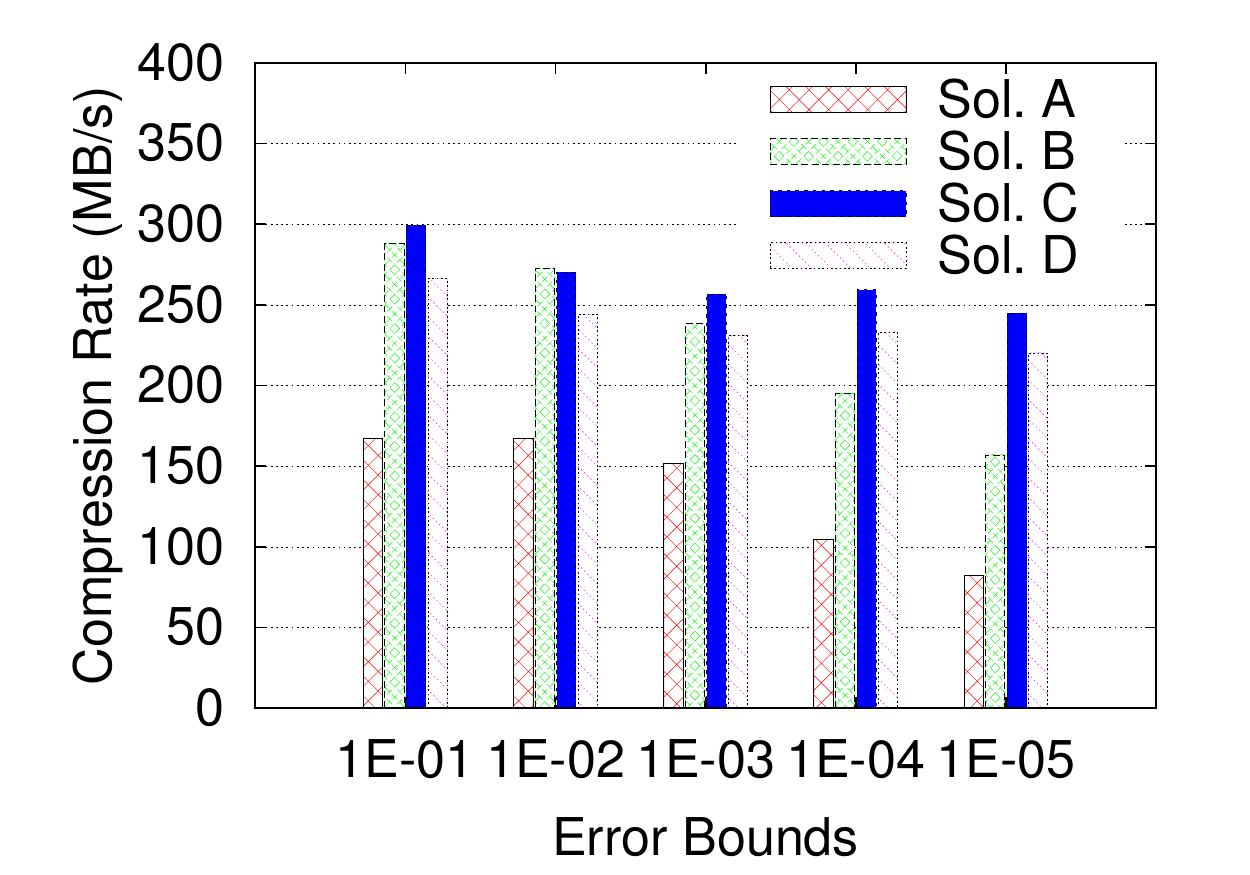}
}
\hspace{-8mm}

\hspace{-7mm}
\subfigure[{qaoa\_36 (Decmpr)}]
{
\includegraphics[scale=.36]{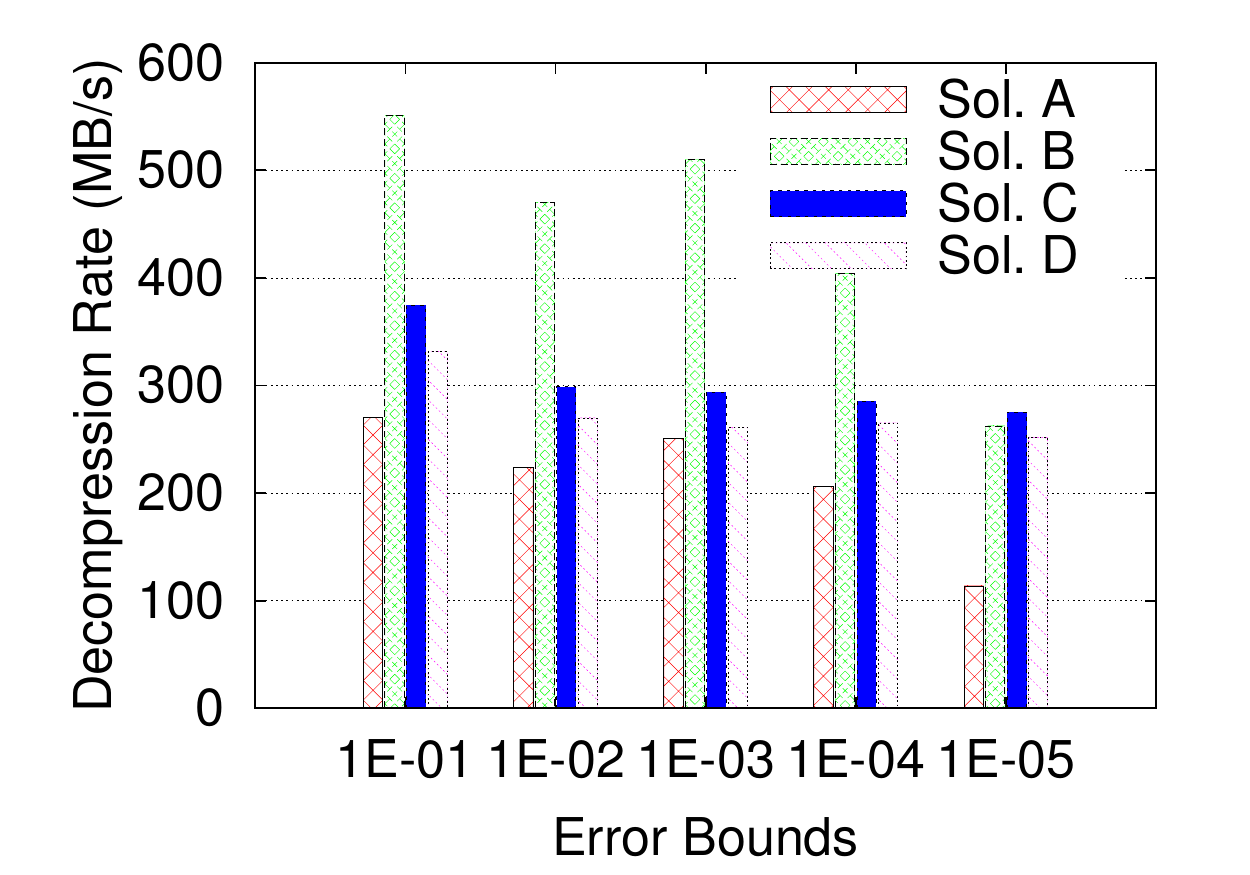}
}
\hspace{-7mm}
\subfigure[{sup\_36 (Decmpr)}]
{
\includegraphics[scale=.36]{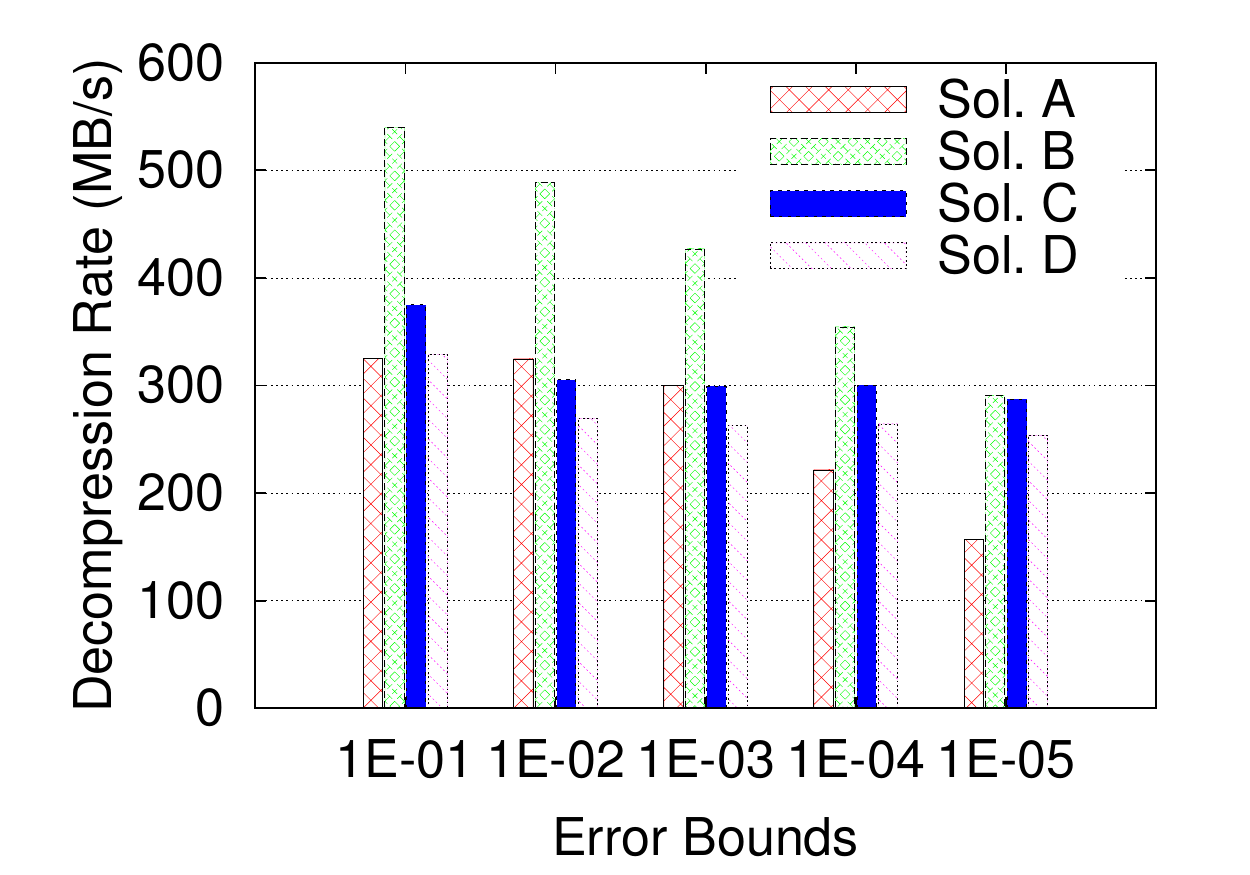}
}
\hspace{-8mm}

\hspace{-7mm}
\subfigure[{qaoa\_36 (Cmpr+Decmpr)}]
{
\includegraphics[scale=.36]{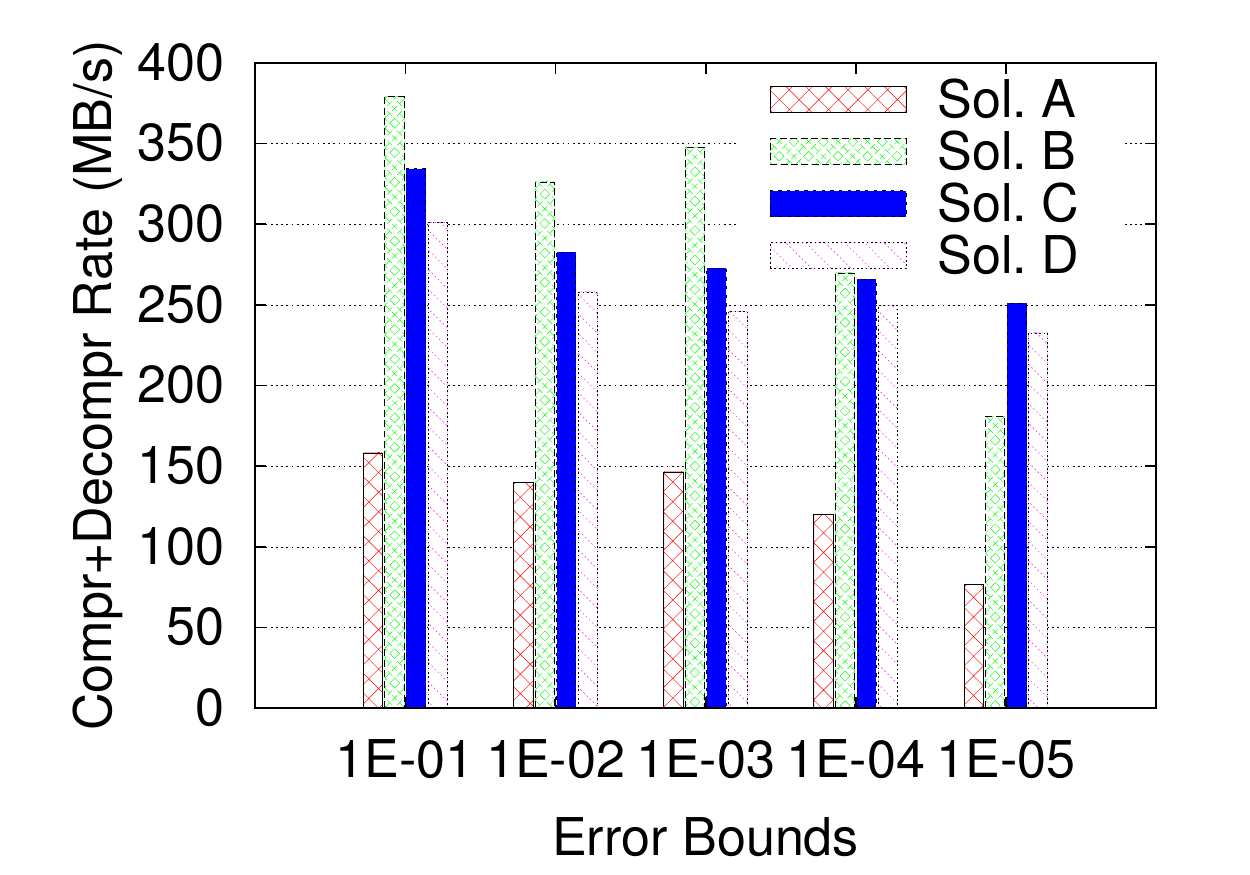}
}
\hspace{-7mm}
\subfigure[{sup\_36 (Cmpr+Decmpr)}]
{
\includegraphics[scale=.36]{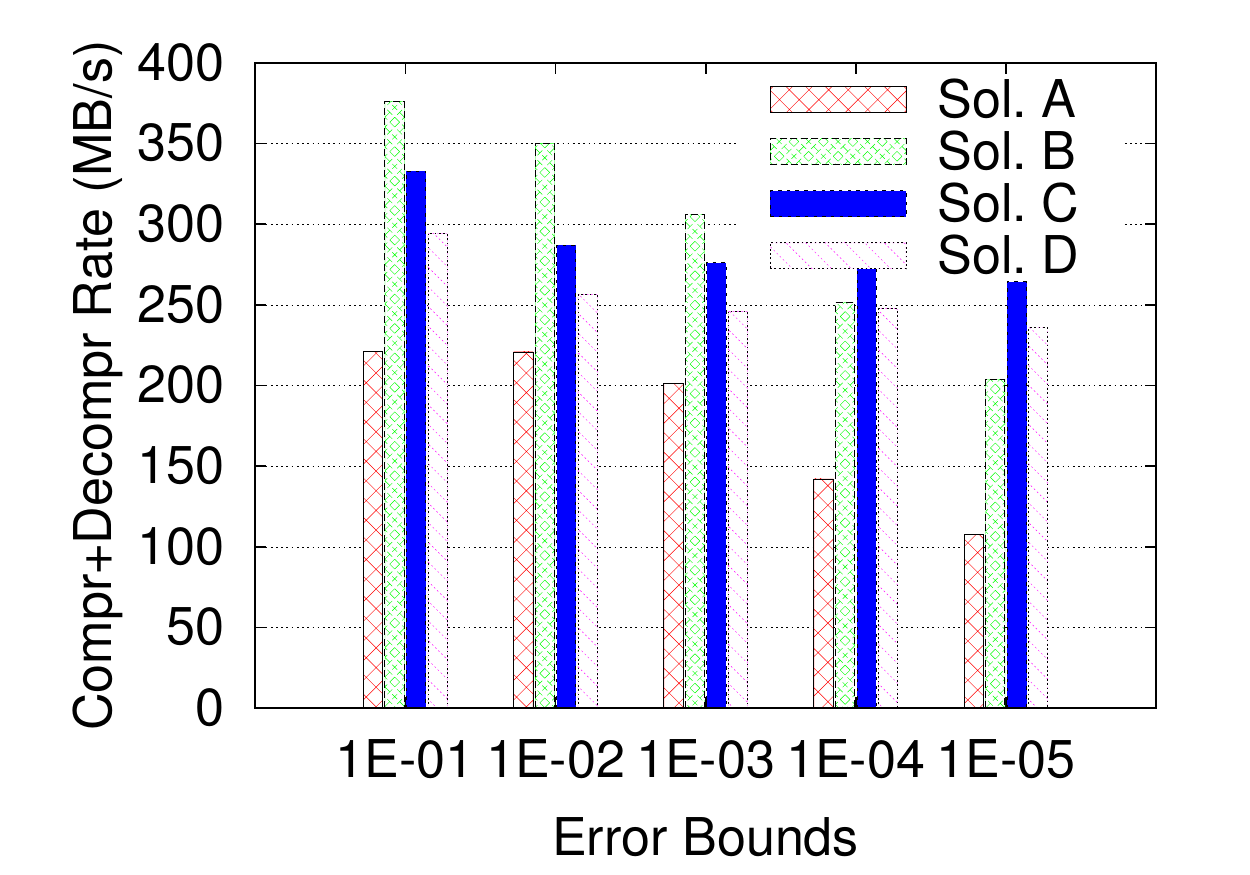}
}
\hspace{-8mm}

\vspace{-4mm}
\caption{Compress/decompression rates (relative error)}
\label{fig:4sol-speed}
\end{figure}

We present the compression and decompression rate in Figure \ref{fig:4sol-speed} (single-core performance). This figure clearly shows that Solutions B, C, and D have much higher compression rates and decompression rates than the classic SZ 2.1 (Solution A) does. The key reason Solution B is faster than Solution A is that it predicts the data in terms of the complex data type, which may get higher prediction accuracy, thus leading to faster encoding thereafter. Moreover, we set a lower maximum number of quantization bins in Solution B such that it keeps a relatively high compression rate in the case with low relative error bounds such as 1E-5. The reason the Solutions C and D run much faster than Solutions A and B do is that they totally remove the three costly steps---prediction, quantization, and Huffman encoding---in the compression. We can also observe that Solution C is slightly faster than Solution D  because of the extra reshuffle step in Solution D.

\begin{figure*}[ht] \centering
\hspace{-8mm}
\subfigure[{qaoa\_36:PWR=1E-1}]
{
\includegraphics[scale=.31]{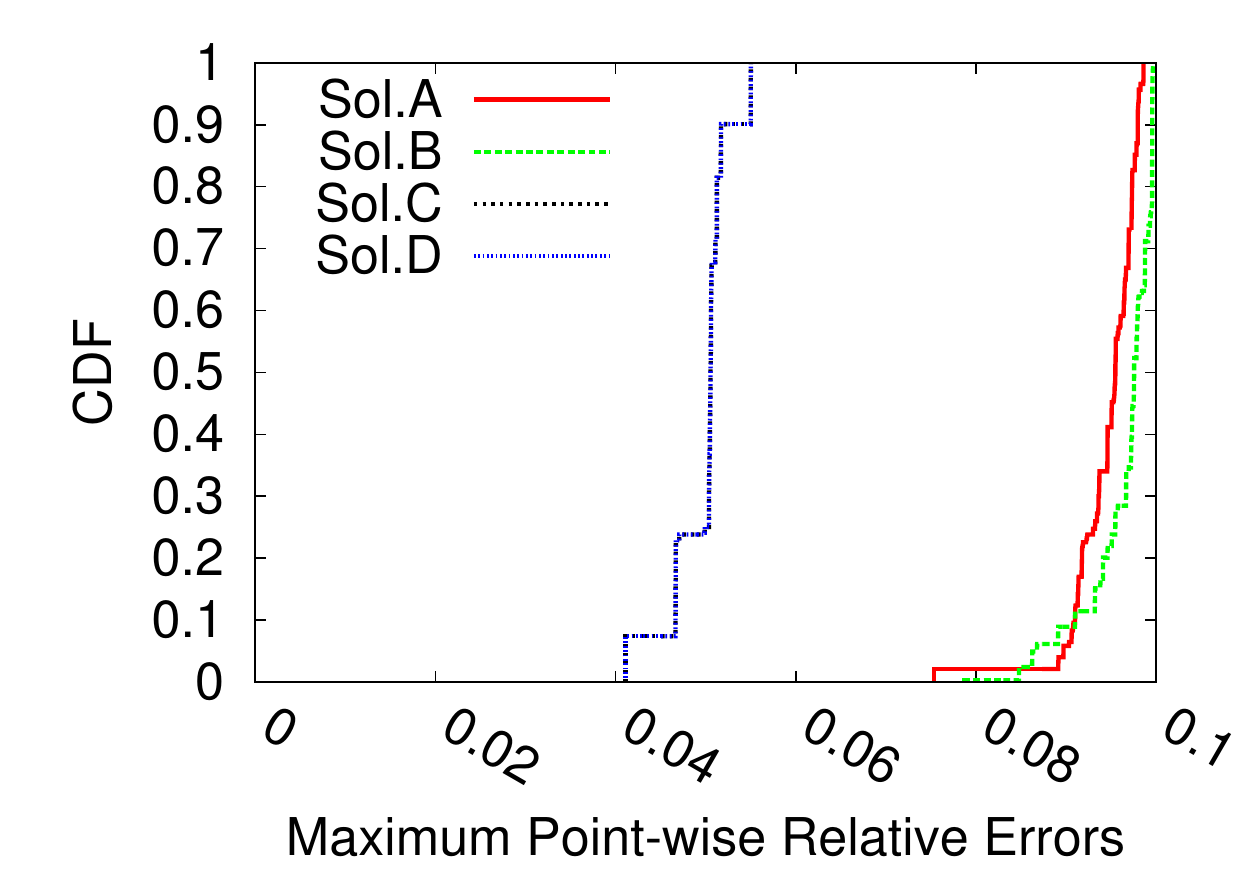}
}
\hspace{-8mm}
\subfigure[{qaoa\_36:PWR=1E-2}]
{
\includegraphics[scale=.31]{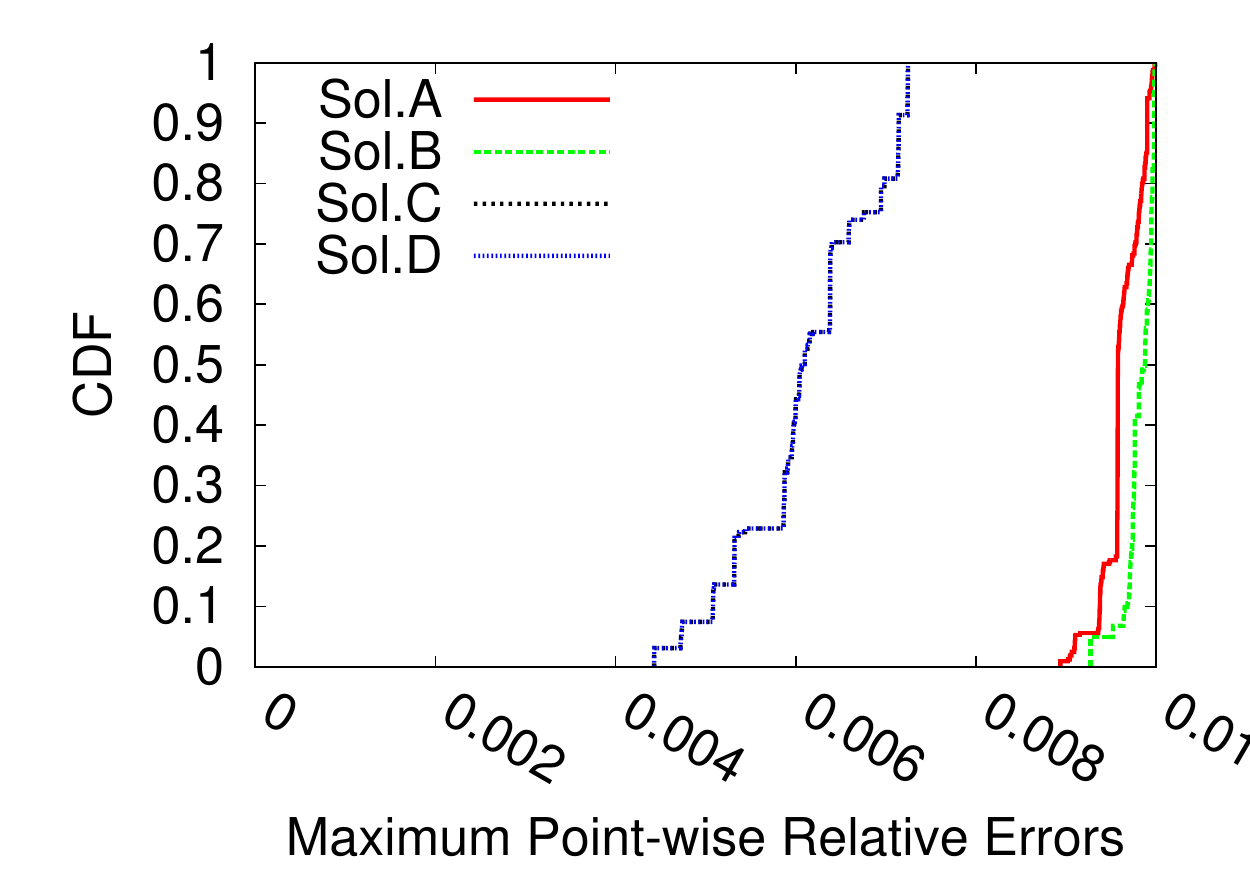}
}
\hspace{-8mm}
\subfigure[{qaoa\_36:PWR=1E-3}]
{
\includegraphics[scale=.31]{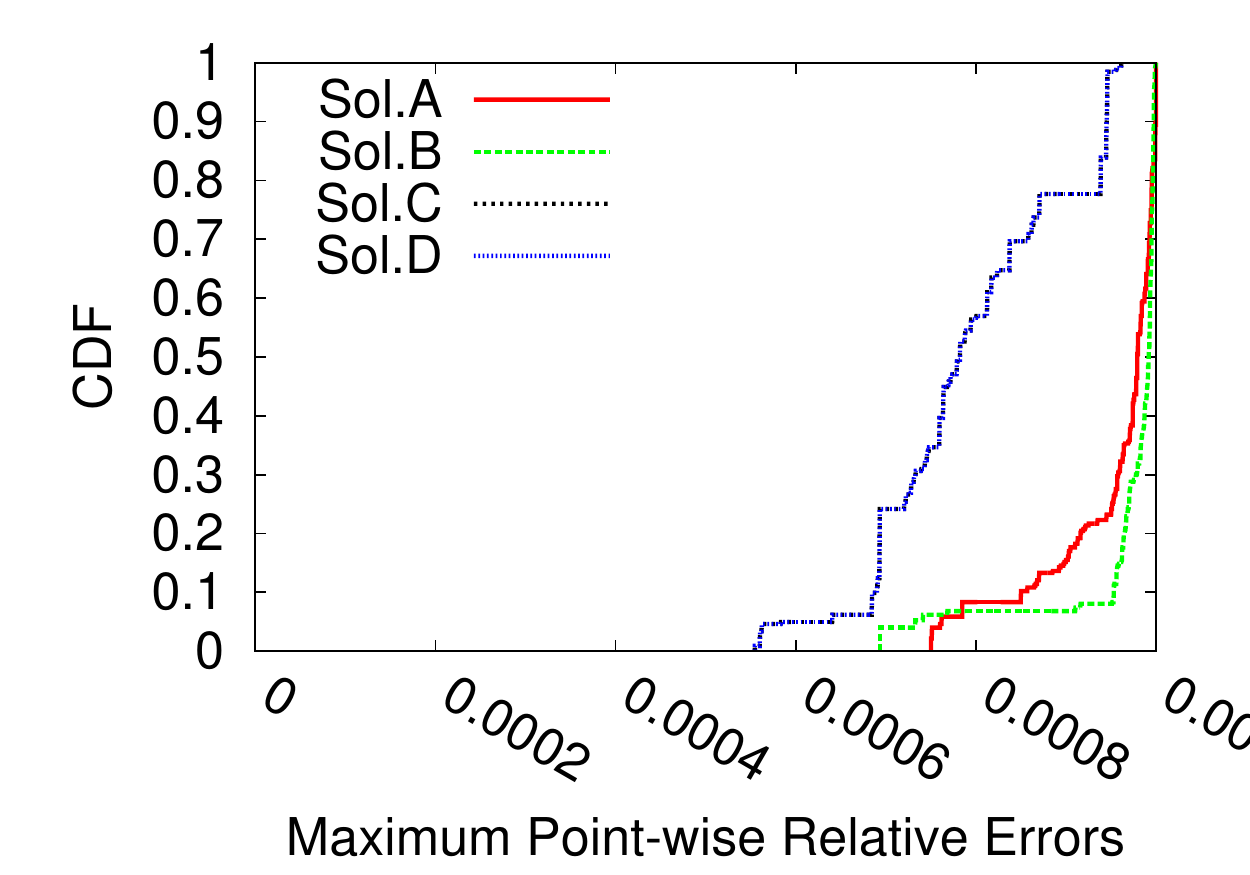}
}
\hspace{-8mm}
\subfigure[{qaoa\_36:PWR=1E-4}]
{
\includegraphics[scale=.31]{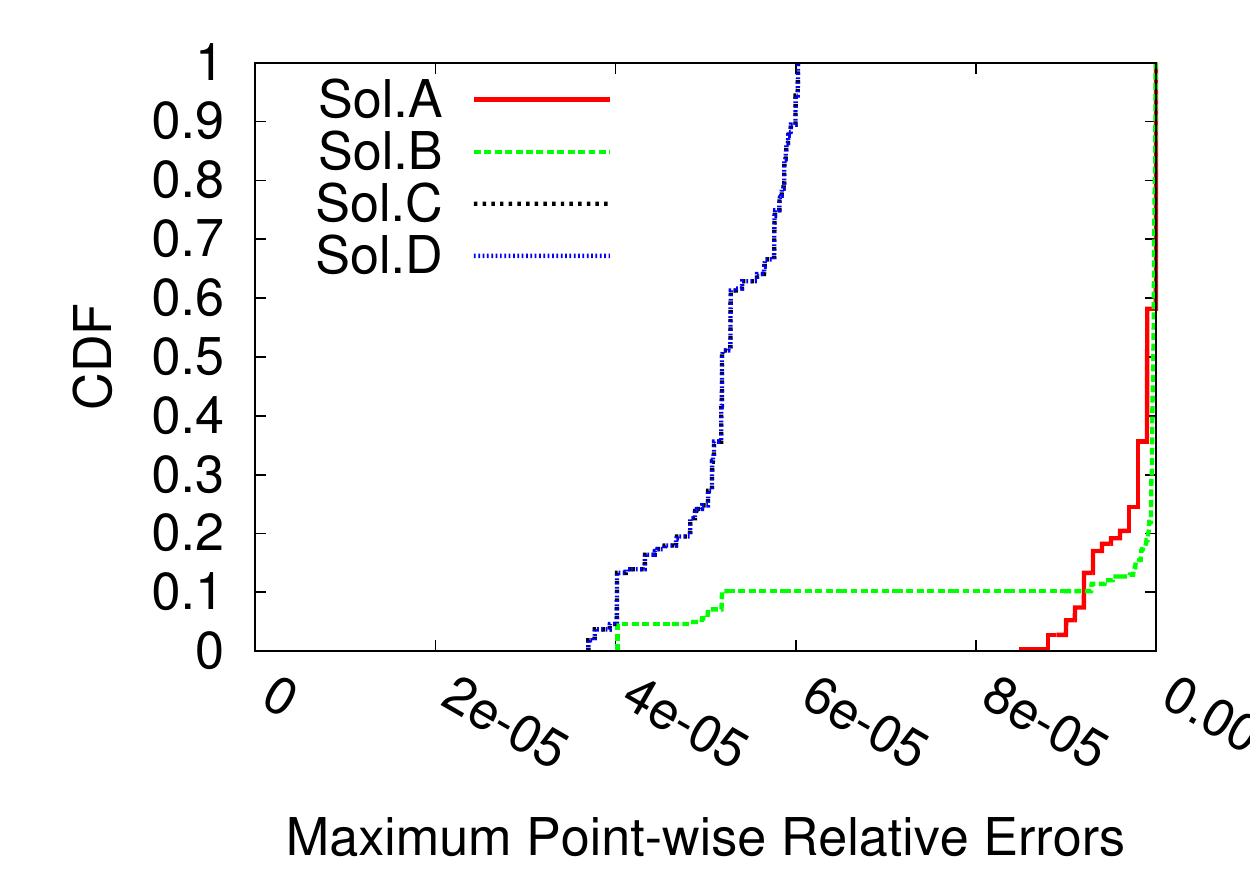}
}
\hspace{-8mm}
\subfigure[{qaoa\_36:PWR=1E-5}]
{
\includegraphics[scale=.31]{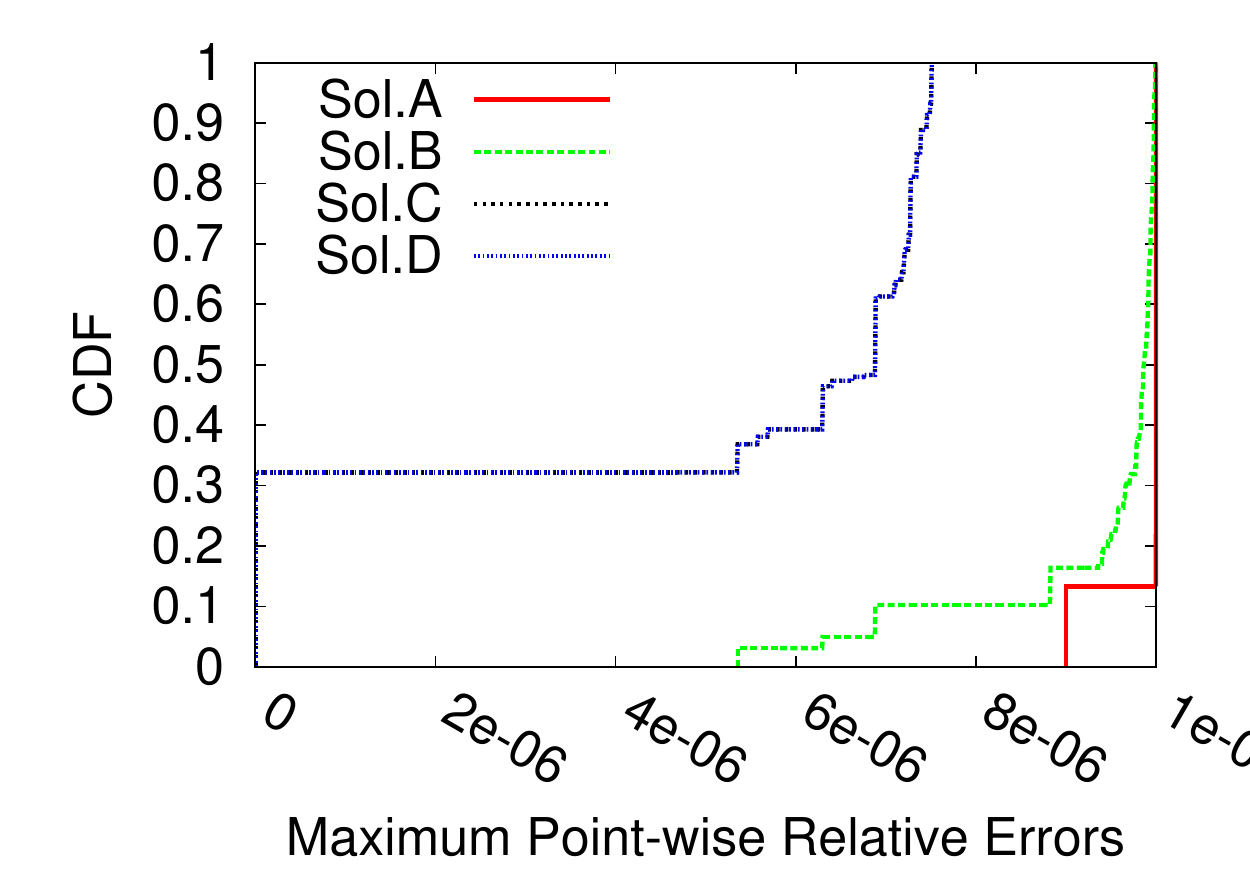}
}
\hspace{-8mm}

\hspace{-8mm}
\subfigure[{sup\_36:PWR=1E-1}]
{
\includegraphics[scale=.31]{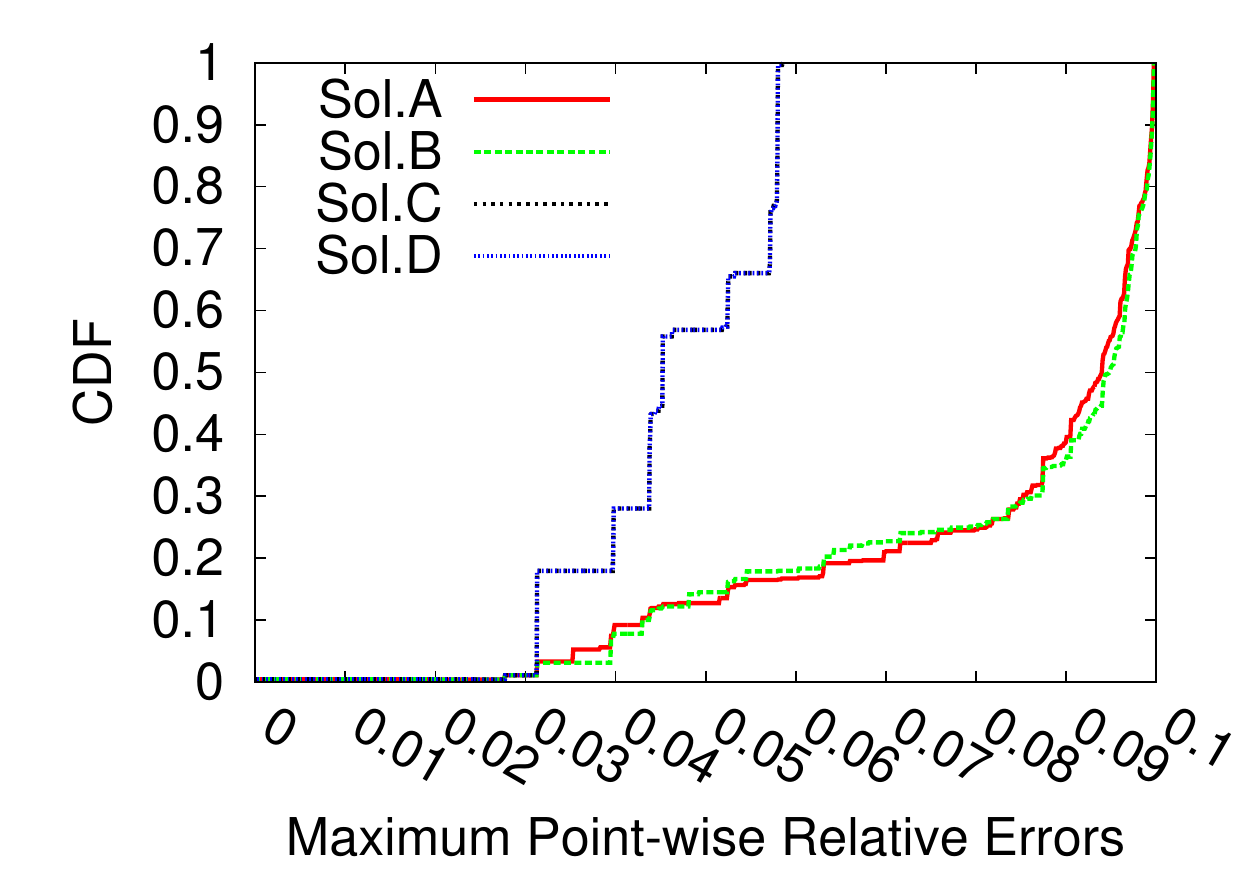}
}
\hspace{-8mm}
\subfigure[{sup\_36:PWR=1E-2}]
{
\includegraphics[scale=.31]{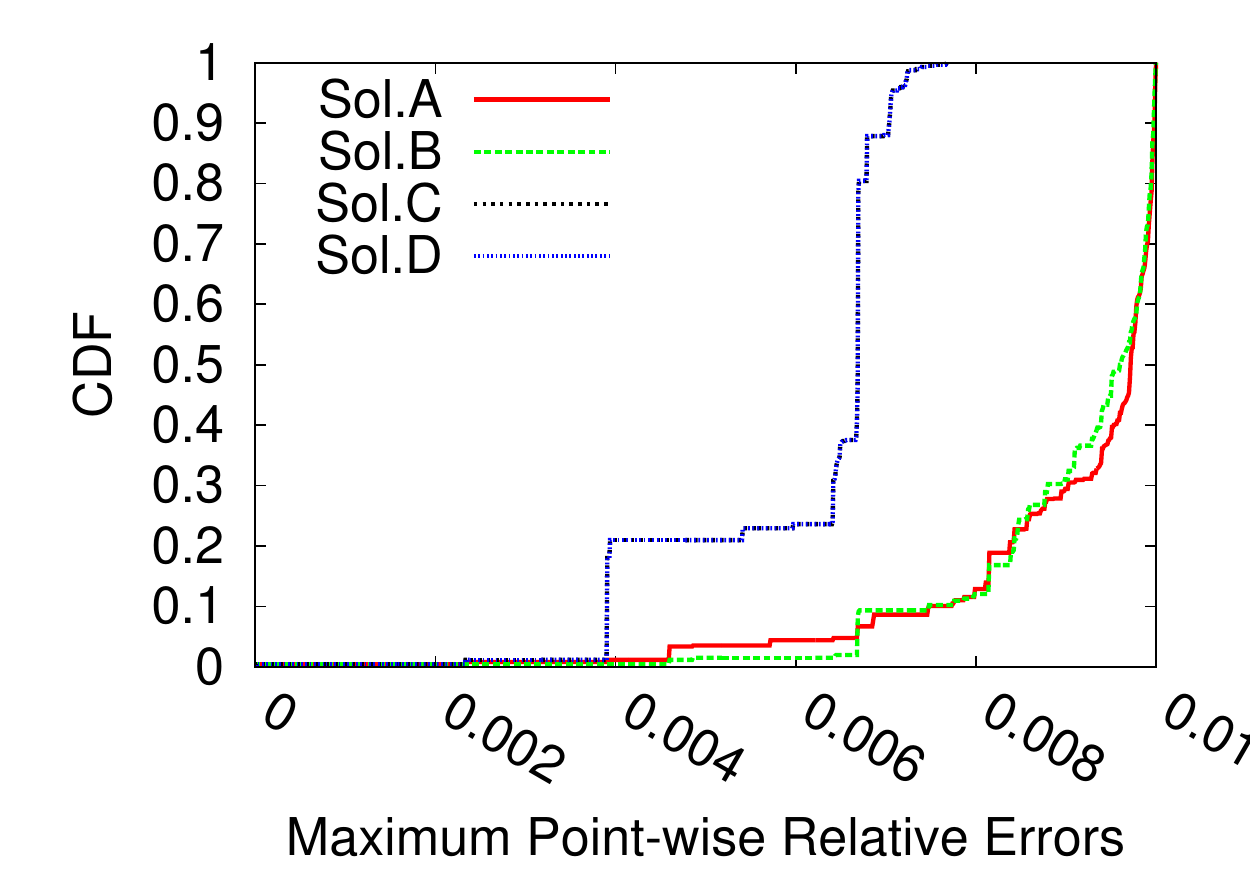}
}
\hspace{-8mm}
\subfigure[{sup\_36:PWR=1E-3}]
{
\includegraphics[scale=.31]{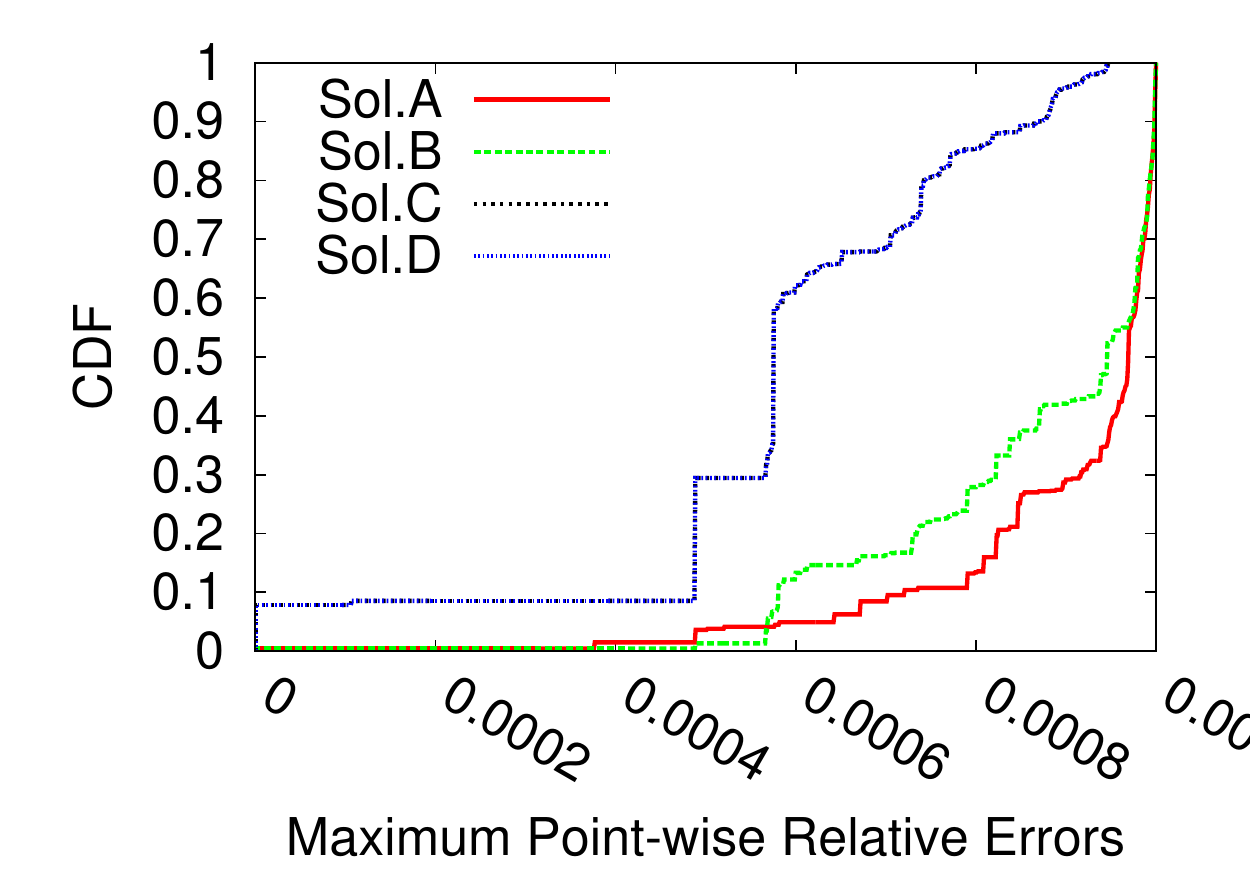}
}
\hspace{-8mm}
\subfigure[{sup\_36:PWR=1E-4}]
{
\includegraphics[scale=.31]{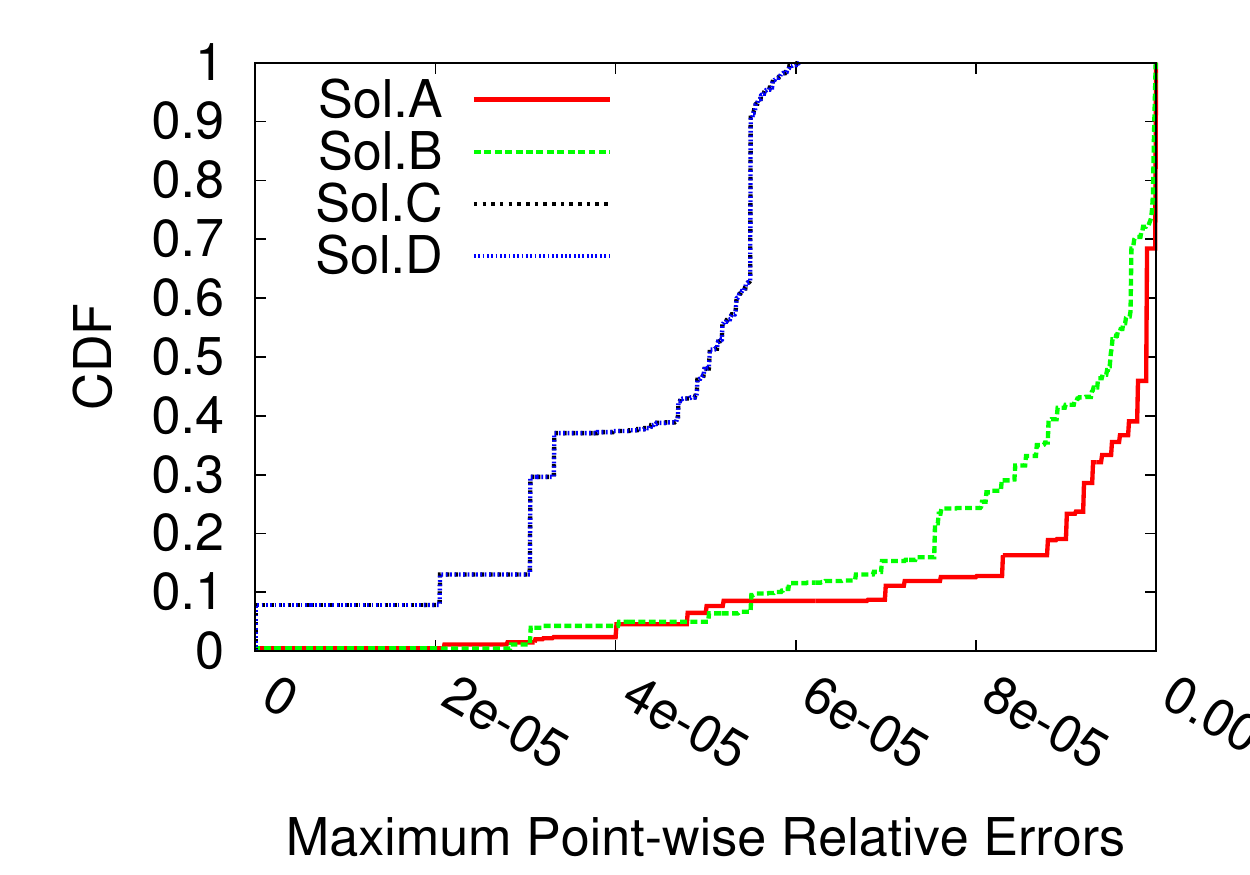}
}
\hspace{-8mm}
\subfigure[{sup\_36:PWR=1E-5}]
{
\includegraphics[scale=.31]{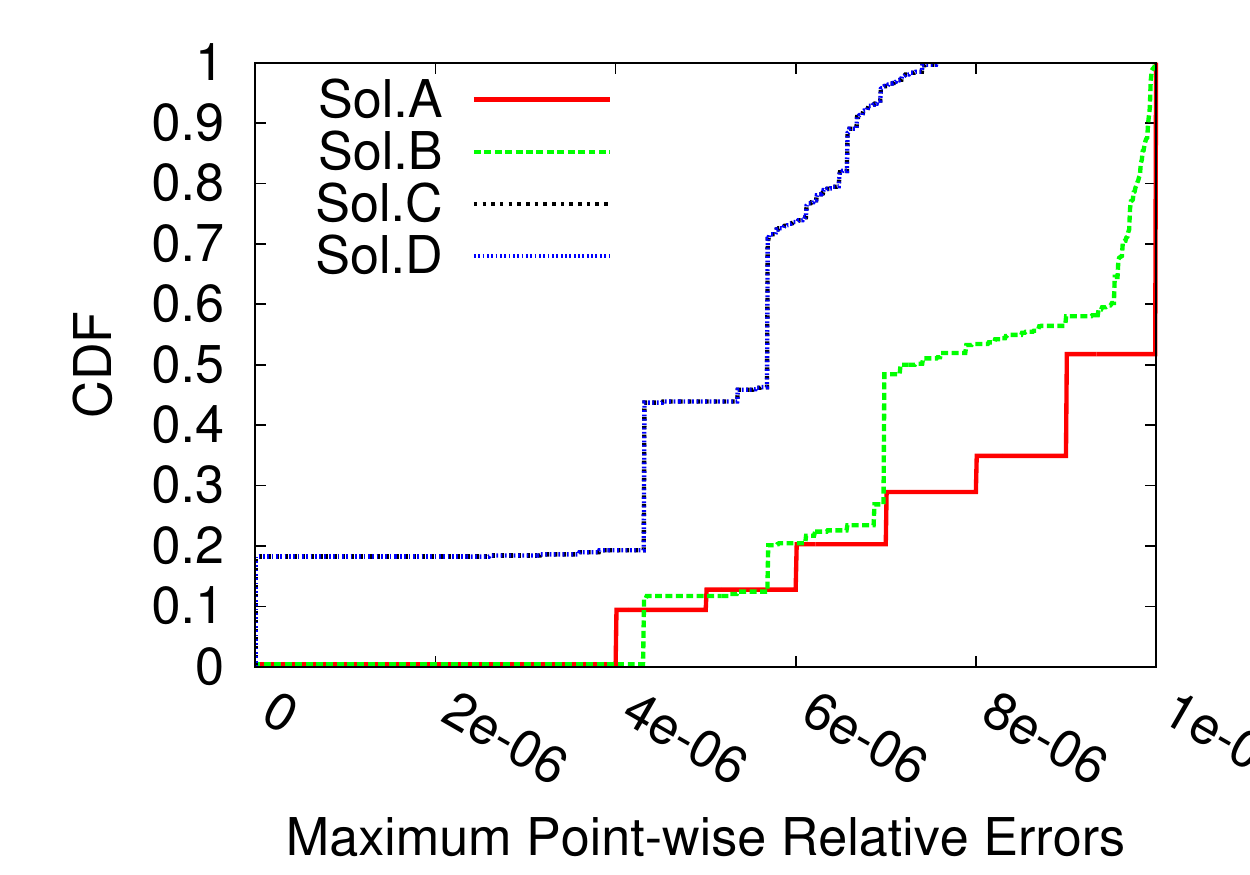}
}
\hspace{-8mm}
\vspace{-4mm}
\caption{Distribution of maximum relative errors on compression of qaoa\_36 and sup\_36 (Solution C and D overlap)}
\label{fig:err-distribution}
\vspace{-4mm}
\end{figure*}



We present in Figure \ref{fig:err-distribution} the distribution of the maximum compression error (pointwise relative errors)  per data block for all four compression solutions on qaoa\_36 and sup\_36, respectively. Similar to the previous evaluations, the experimental dataset comes from one rank involving hundreds of data blocks (each with 1,048,576 complex-type data points, bringing a total of 16 MB per block). 

From Figure \ref{fig:err-distribution} we observe that all four solutions respect the compression errors  well in terms of different pointwise relative error bounds (1E-1$\sim$1E-5). We note that the error distribution curves of Solutions C and  D overlap, which is explained as follows. In fact, they both avoid the data prediction and quantization step such that the compression errors have nothing to do with the order of the data points. The only difference between them is the extra reshuffle step in Solution D. Thus they have exactly the same compression errors in each block. 

\begin{figure}[ht] \centering
\subfigure[{Prediction and quantization in solution A/B}]
{
\includegraphics[scale=.44]{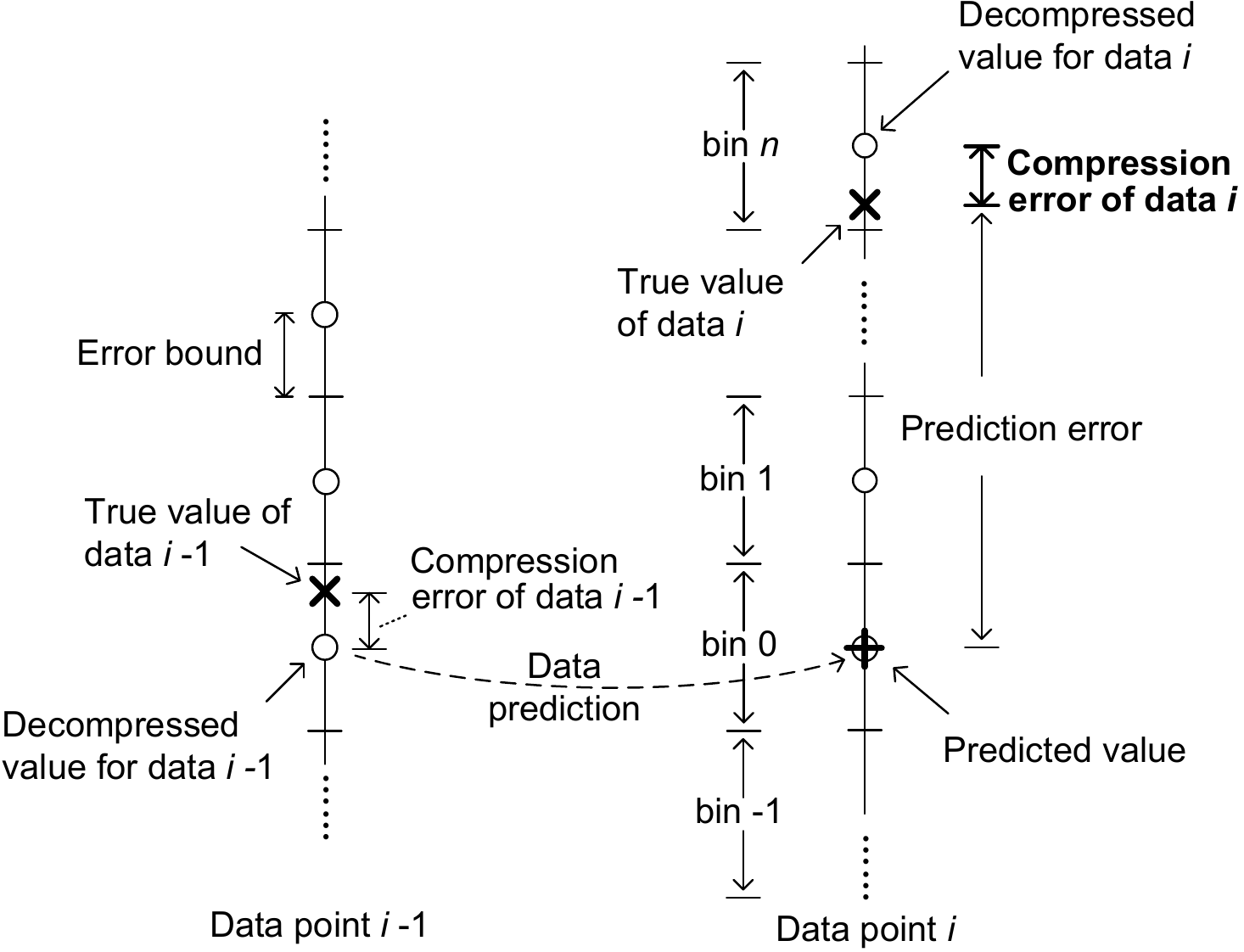}
}
\subfigure[{Discrete relative errors when truncating  bits in solution C/D}]
{
\includegraphics[scale=.4]{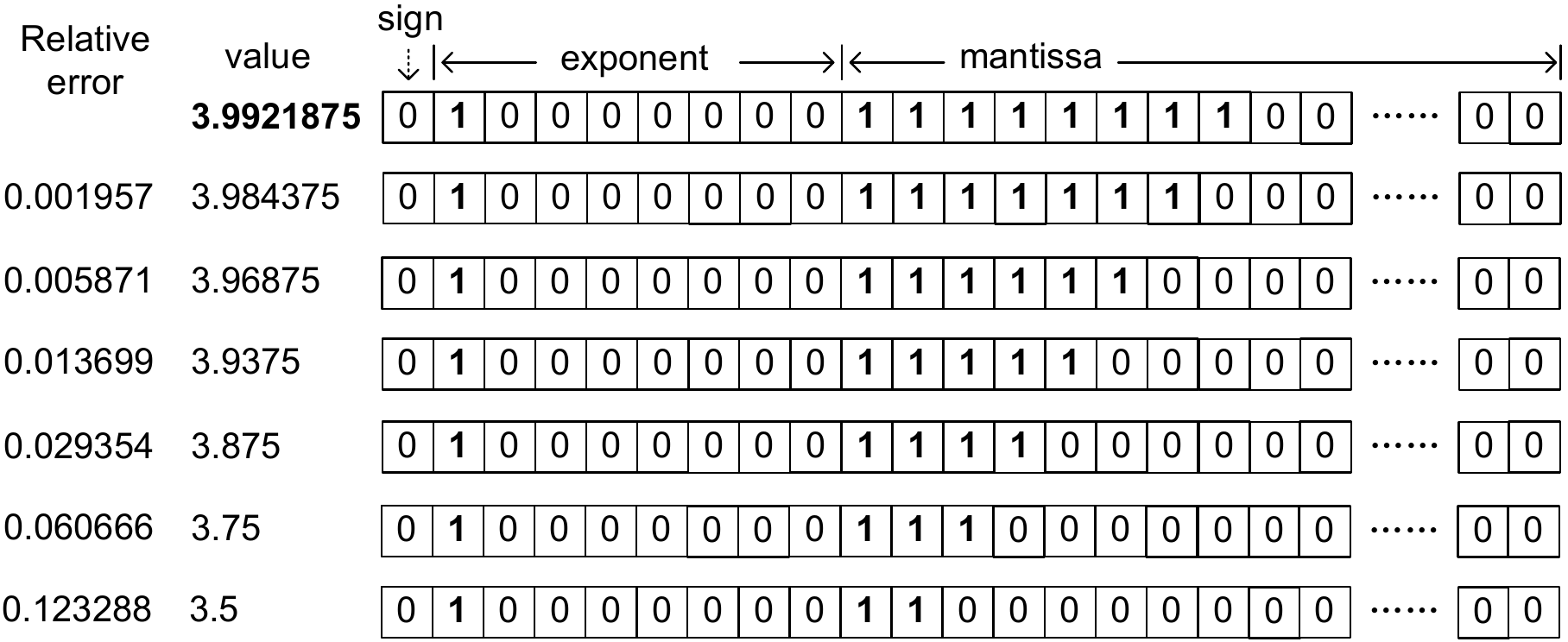}
}

\vspace{-4mm}
\caption{Illustrating why Solution C/D leads to lower compression errors than Solution A/B}
\label{fig:expl-cmpr-err}
\end{figure}

Moreover, we note that Solutions C and D exhibit much lower compression errors than do Solutions A and B in general. The reason is  as follows.  As illustrated in Figure \ref{fig:expl-cmpr-err} (a), Solutions A and B both adopt the data prediction and linear-scaling quantization \cite{tao2017significantly}. Specifically, they predict each data point and then approximate its value by the error-bound-quantized distance \cite{tao2017significantly} between the true raw value and the predicted value. Note that the compression errors are determined by the difference between the decompressed values and the true values (as shown in Figure \ref{fig:expl-cmpr-err} (a)). Hence, if the error bound is relatively small, the quantization bin size would be small too, then the true values would be located at a rather random postion in a quantization bin because of the likely large distance between the predicted value and true value, thus leading to a uniform distribution. By comparison, Solutions C and D calculate the significant bit-planes based on user-required pointwise relative errors, and each bit-plane spans a certain value range. As illustrated in Figure \ref{fig:expl-cmpr-err} (b) (with a single-precision value 3.9921875 as an example), truncating different bit planes will lead to discrete decompressed values and relative errors. Suppose the relative error bound is set to 0.01, then we need to keep 15 leading bits and the decompressed value would be 3.96875 with a relative error of 0.005871, which is actually lower than the error bound 0.01. That is, the compression errors of the solutions C and D are generally somewhat lower than the desired error bound. 

\begin{figure}[ht] \centering
\vspace{-2mm}
\hspace{-7mm}
\subfigure[{qaoa\_36}]
{
\includegraphics[scale=.36]{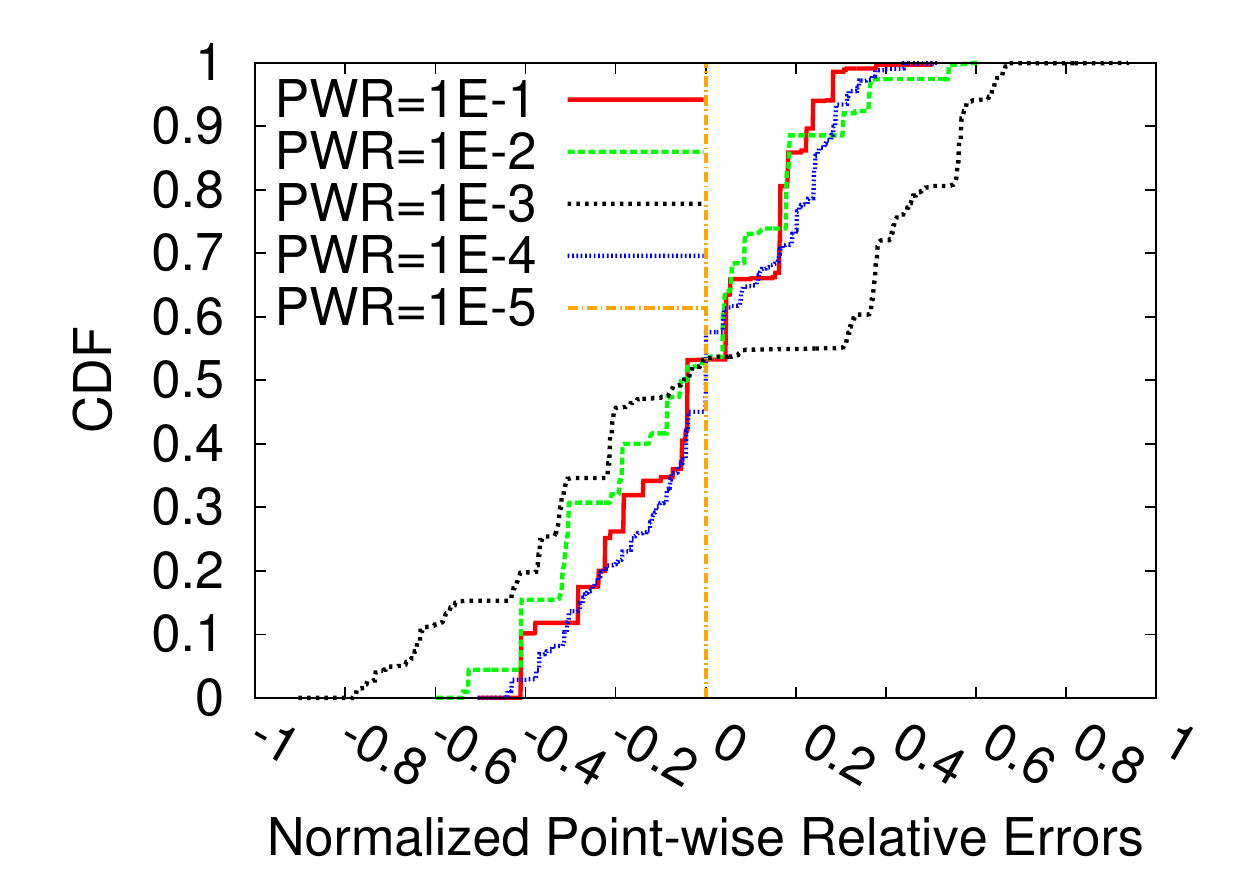}
}
\hspace{-7mm}
\subfigure[{sup\_36}]
{
\includegraphics[scale=.36]{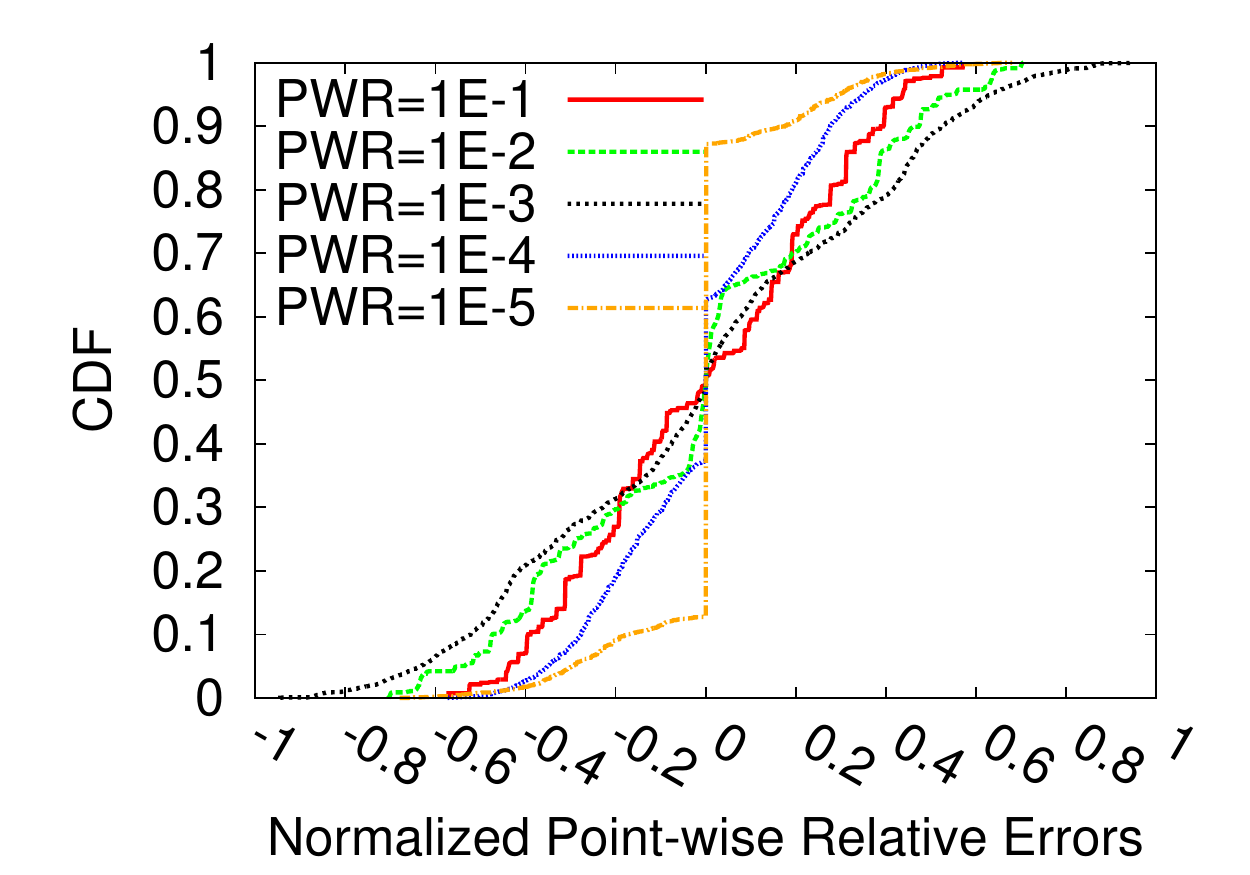}
}
\hspace{-8mm}

\vspace{-4mm}
\caption{Distribution of normalized compression errors (Solution C)}
\label{fig:normalized-err-distribution}
\vspace{-2mm}
\end{figure}

Such overpreservation of error can be observed more clearly by plotting the distribution of normalized compression errors compared with the error bounds for Solution C, as presented in Figure \ref{fig:normalized-err-distribution}. We plot the cumulative distribution function (CDF) of the normalized pointwise relative errors compared with the corresponding error bounds for one random block of data, because too many data points are involved in the whole simulation and other blocks actually exhibit similar error distributions. In the figure we can observe that (1) all compression errors are indeed confined within the required relative error bound; (2) the compression errors follow a  uniform distribution; and (3) most of the compression errors are actually much lower than the required error bound, bringing an extra benefit to the control of data distortion in the simulation.

Moreover, the solution C leads to non-correlated compression errors (point-wise relative errors) for the quantum circuit simulation datasets. The reason is that the relative errors are determined by the high-order bit-values in the truncated insignificant bit-planes. Since the quantum circuit simulation data exhibit a rather high randomness (as presented in Figure \ref{fig:illus-data}), the truncation errors are supposed to be fairly random as well. We confirm this point by calculating the lag-1 autocorrelation coefficients of the compression errors. The autocorrelation value is always in [-1,1]; the closer to zero, the higher level of non-autocorrelation is. Our evaluation shows that the autocorrelation value often ranges in [-1E-4,1E-4] if a large majority of original raw data are non-zeros in the dataset, testifying the high non-correlation feature of the solution C. 

Based on this analysis, we can conclude that  Solution C is a fairly good tradeoff, which can obtain a high compression ratio, low compression time and decompression time, and low compression errors with non-correlation feature. Accordingly, Solution C is our final error-bounded lossy compressor to be used in our experiments.

\section{Evaluation}\label{evaluation}


\subsection{Experimental Setup}
For multinode evaluation, we performed our simulation on the Theta supercomputer at  Argonne National Laboratory. Theta consists of 4,392 nodes, each node containing a 64-core Intel\textregistered  Xeon Phi{\tiny\textsuperscript{TM}} processor 7230 with 16 gigabytes  of high-bandwidth in-package memory (MCDRAM) and 192 GB of DDR4 RAM \cite{parker2017early}. The bandwidth of the MCDRAM is 400GB/s, and the average latency is 154 ns \cite{asai2016mcdram}. On Theta, there are wall-time limits for each jobs with different numbers of nodes. The wall-time limits are 3, 6, and 24 hours for jobs of 128 nodes, 256 nodes, and 1,024 nodes, respectively.  For jobs running exceeding wall-time limits, we use our checkpoint design, described in Section~\ref{sec:checkpoint}, to complete the simulation.

For single-node experiments, we run our simulation on a single 64-core Intel\textregistered Xeon Phi{\tiny\textsuperscript{TM}} processor 7210 (KNL) in the Joint Laboratory for System Evaluation (JLSE) at Argonne. There is no wall-time limit on the system.
 
In this evaluation, we set 128 ranks per node. As described in Section~\ref{overview}, the state vector on each rank is divided into multiple blocks, each containing 1,048,576 amplitudes, which means each block is 16 MB. Since each rank decompresses at most 2 blocks on MCDRAM at the same time, we need to  allocate $128 \times 2 \times 16 MB = 4 GB$ of memory on MCDRAM. Thus, we configure the memory mode by using \emph{\texttt{-{}-}attrs mcdram=equal} to operate 50\% MCDRAM as flat memory and the other 50\% as cache \cite{jeffers2013intel}.

\subsection{Scalability}
\begin{figure}[ht]
\centering
\includegraphics[scale=.4]{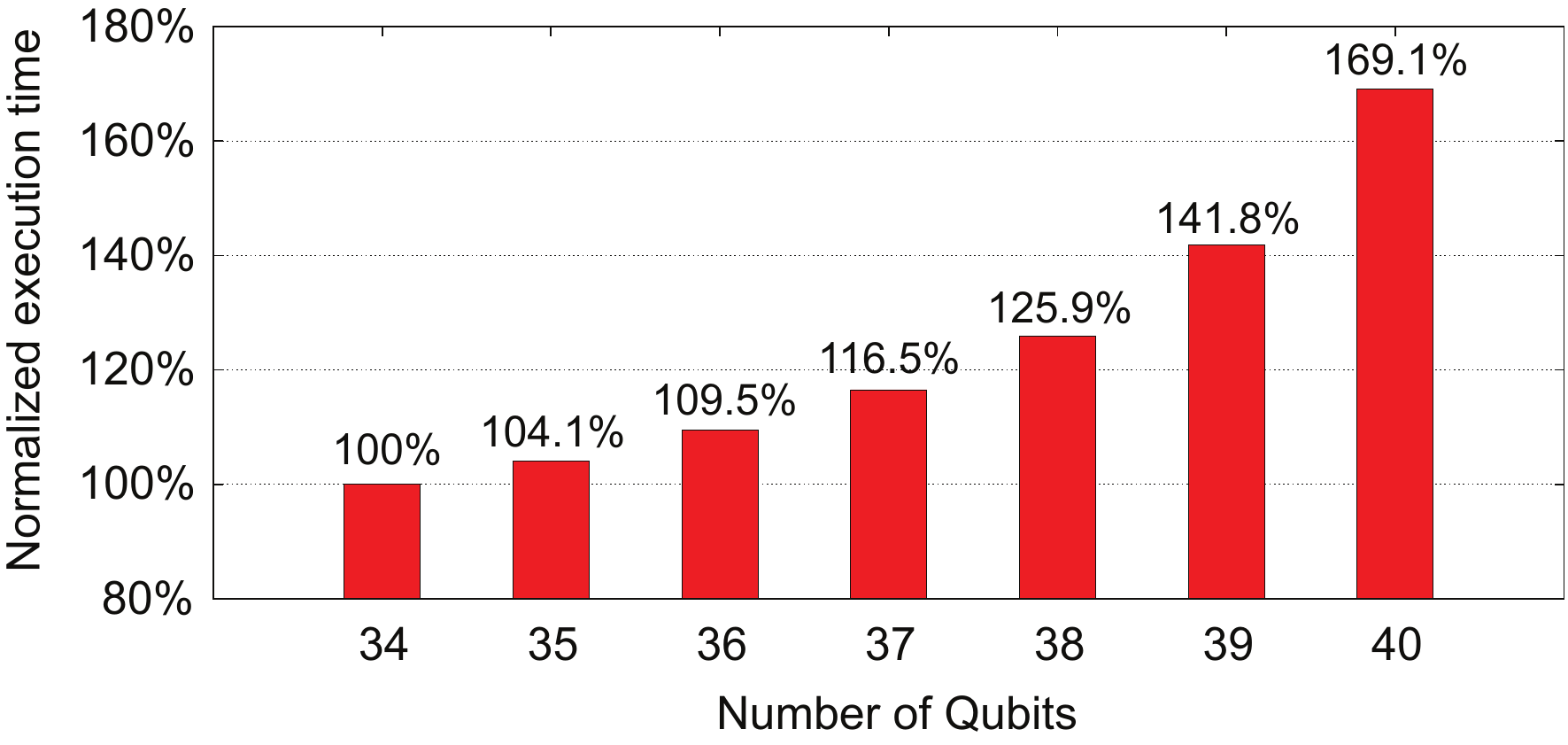}
\vspace{-3mm}
\caption{Normalized execution time for running various sizes of simulations on a single node.}
\label{fig:scale_qubits}
\end{figure}

\begin{figure}[ht]
\centering
\includegraphics[scale=.4]{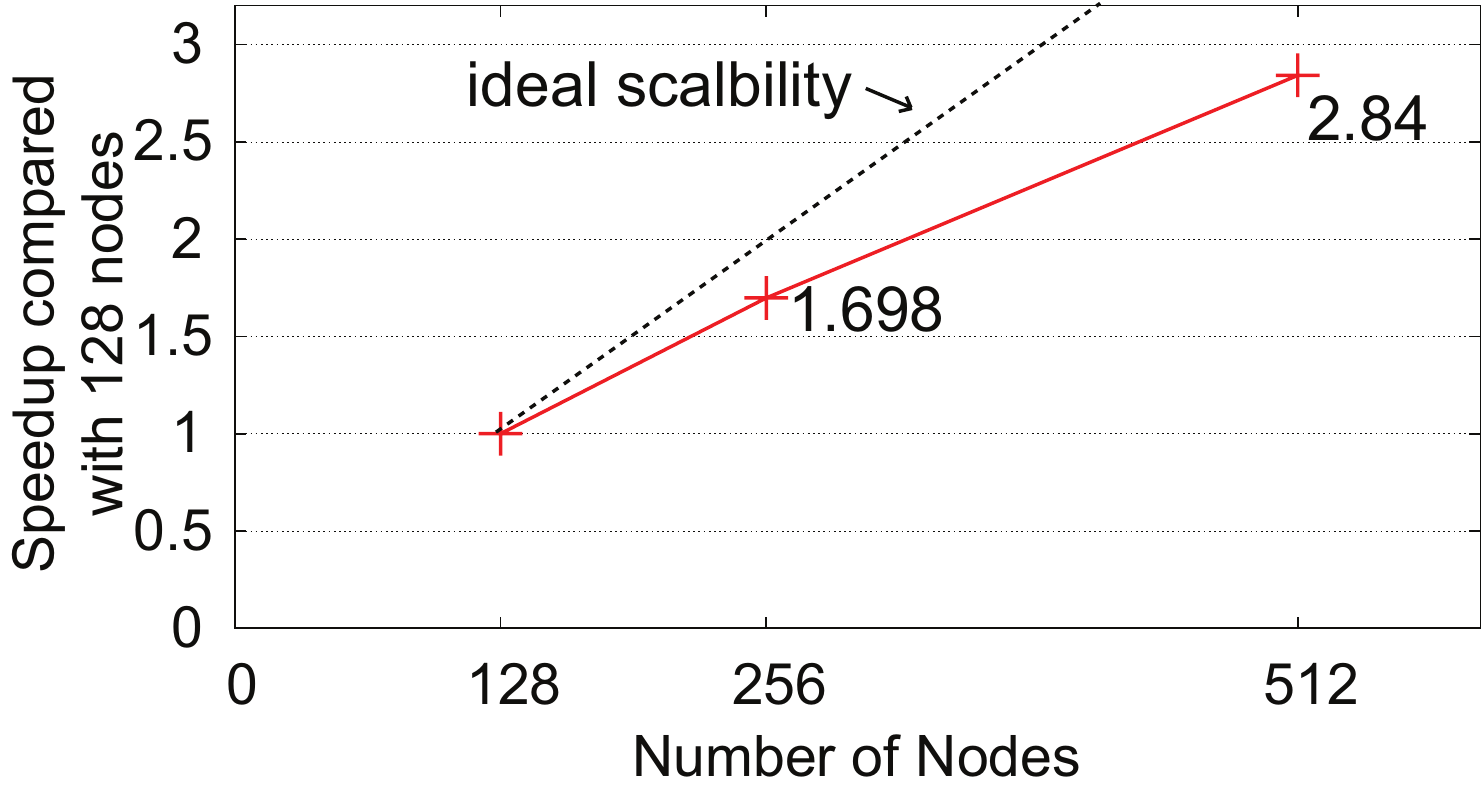}
\vspace{-4mm}
\caption{Scaling behavior of applying Hadamard gates on a 51-qubit system on Theta.}
\label{fig:scale_nodes}
\vspace{-4mm}
\end{figure}

We evaluate the scaling behavior by using a basic program that applies a Hadamard gate on each qubit. Figure~\ref{fig:scale_qubits} shows the normalized execution time for running different sizes of simulations on a single node. The scaling of our simulation technique for 51-qubit quantum circuit running on different numbers of nodes is depicted in Figure~\ref{fig:scale_nodes}. For simulations of real applications that requires more time in compression, decompression, and computation, the scaling is expected to be better.

\subsection{Benchmarks}
To show the results against real applications, we choose multiple important quantum applications as our benchmarks. The benchmarks are selected to have different program characteristics in order to show that our simulation could work with arbitrary quantum circuits. Our approach performs well for both shallow-circuit and deep-circuit applications. The initial state is $\ket{0}^{\otimes n}$. The programs we use for our benchmarking include the following.
\begin{itemize}
\item Grover: Grover's search algorithm is for database search, and it leads to significant speedups compared with classical search algorithms \cite{grover1996fast, javadiabhari2015scaffcc}. Our benchmark uses Grover's search algorithm to find the square root number \cite{javadiabhari2015scaffcc}, and thus the oracle consists of X and Toffoli gates.
\item Random circuit sampling: The circuit is proposed by Google to show the quantum supremacy, and we follow the rules to construct the random circuits \cite{boixo2018characterizing}. Since our approach is not optimized for the circuits, we do not plan to run the random circuits with many layers. Thus, the circuit depth is 11 in our experiments. 
\item QAOA: The quantum approximate optimization algorithm is a hybrid quantum-classical variational algorithm. Our benchmark uses QAOA to solve MAXCUT on a random 4 regular graph problem \cite{farhi2014quantum}. QAOA is an important circuit because it is one of the most promising quantum algorithms in the NISQ era \cite{preskill2018quantum}.
\item QFT: This is the quantum circuit for quantum Fourier transform \cite{javadiabhari2015scaffcc}, which is an important function in many quantum algorithms (Shor's algorithm \cite{shor1999polynomial}, phase estimation algorithm \cite{cleve1998quantum}, and the algorithm for hidden subgroup problem \cite{jozsa2001quantum}), and this is a deep circuit. We randomly apply X gate to the initial state as the input for the QFT in our experiments.
\vspace{-2mm}
\end{itemize}

\begin{table*}[ht]
\centering
  \caption{\textbf{Experimental results.} The first row shows the memory requirements of the simulations without our techniques. The ratios of our system memory sizes to the required memory sizes are presented in the fourth row. The fifth row provides the simulation time and the breakdown. The minimum compression ratio during the simulation is shown in the last row.}
  \vspace{-3mm}
  \small
  \begin{tabular}{| l | c c c | c c  c c | c c c  | c |}
    \hline
    Benchmark & \multicolumn{3}{c|}{Grover} & \multicolumn{4}{c|}{Random Circuit Sampling} & \multicolumn{3}{c|}{QAOA} & QFT\\ \hline\hline
    Number of Qubits & 61 & 59 & 47 & $5\times9$&  $6\times7$ & $6\times6$ & $7\times5$ & 45 & 43 & 42 & 36\\
    (Memory Requirement) & (32 EB) & (8 EB) & (2 PB) & (512 TB)  & (64 TB) & (1 TB) & (512 GB) & (512TB) & (128 TB) & (64 TB) & (1 TB)\\ \hline
    Number of Gates & 314 & 310 & 305 & 227  & 261 & 165 & 208  & 394 & 344 & 336  & 3258\\ \hline
    Number of Nodes & 4096 & 4096 & 128 & 1024  & 128 & 1 & 1  & 1024 & 256 & 128 & 1\\ \hline
    Total System  Memory & 768 TB & 768 TB & 24 TB & 192 TB &  24 TB & 192 GB & 192 GB  & 192TB & 48 TB & 24 TB  & 192 GB\\ 
    (Sys Mem / Req.) & (0.002\%) & (0.009\%) & (1.17\%) & (37.5\%) & (37.5\%) & (18.75\%) & (37.5\%)  &  (37.5\%) & (37.5\%) & (37.5\%)  & (18.75\%)\\\hline
    Total Time  (Hour) & 8.14 & 3.48    & 0.49 & 4.87  & 8.64 & 7.96 & 6.23 &  13.34 & 5.83 & 8.65 & 78.98\\ 
    Compression Time  & 1.87\% & 4.59\%  & 2.04\%  & 55.79\%   & 40.26\% & 59.10\% & 58.57\% &  50.66\% & 44.97\% & 41.02\%& 57.86\%\\ 
    Decompression Time & 1.87\% & 3.73\%  & 4.08\%  & 31.47\%  & 22.19\% & 33.78\% & 30.59\% &  26.46\% & 27.64\% & 25.52\%& 37.68\%\\ 
    Communication Time & 32.7\% & 20.98\% & 36.73\% & 0.12\%  & 0.57\% & 0.02\% & 0.03\% &  3.03\% & 0.22\% & 0.23\%&  2.56\%\\ 
    Computation Time & 63.47\% & 70.70\% & 57.15\% & 12.60\%  & 36.97\% & 7.08\% & 10.8\% &  19.84\% & 27.16\% & 33.22\%&  1.9\%\\ \hline
    Time per Gate (Sec) & 93.34 & 40.49 & 5.78 & 64.69 &  119.22 & 173.65 & 107.86 &   121.91 & 61.02 & 92.64 &  87.27\\ \hline
    Simulation Fidelity & 0.996 &0.996 & 1 & 0.987  & 0.993 & 0.933 & 0.985 & 0.895 & 0.999 & 0.999 &  0.962\\ \hline
    Compression Ratio & $7.39\times10^4$ &$8.26\times10^4$ & $1.06\times10^4$ & 6.03 & 9.40 & 8.16 & 10.05 & 5.38 & 4.85 & 9.25 &  21.34\\ \hline
  \end{tabular}

\label{tab:results}
  \vspace{-3mm}
\end{table*}

\subsection{Experimental Results}

We present our main results in Table~\ref{tab:results}. We run benchmarks with the total system memory capacities much less than the theoretical memory requirements to manifest the strength of our approach.

Our first benchmark application is Grover's search algorithm with 47, 59 and 61 qubits. Previously, 61-qubit simulations of general quantum applications were  thought to be impossible because of infeasible memory requirements. Using our methods, the state vectors can be compressed with high compression ratios because the amplitude values of the state vectors of this application are similar and regular, and hence our approach can use only 768 TB to successfully perform the simulation of 61-qubit Grover's search algorithm, which theoretically requires 32 EB to execute the full-state simulation. We also use our technique to complete the 47-qubit Grover's search algorithm simulation using 24 TB instead of 2 PB, the memory requirement without our technique. Next, we present the simulation of Google's quantum supremacy random circuits. Since our method exploits non-uniformity and structure in quantum computations, it does not work as well on random circuits. If we run many layers of the circuits, we have to drop the simulation fidelity to meet the available memory capacity. Thus, we present the simulation results of depth of 11 random circuits. Although our simulation technique is not designed specifically for the supremacy circuits, our approach can simulate the circuit with 42 qubits by using 48 TB, and simulate the circuit with 45 qubits by using 192 TB. The next benchmark is QAOA. Since QAOA is an important quantum application, it is critical to the simulation of QAOA circuits with limited memory capacity. In fact, we can get higher compression ratios and still get successful results because QAOA is robust to low-fidelity. Finally, the results with QFT circuits show that our techniques are effective for simulating high-depth circuits and compress the state vectors with more than 21X compression ratios, with a fidelity of 0.962. 

For the quantum applications we choose, we show that our technique can simulate the circuits with high fidelity. Among the selected benchmarks, random circuits use more entanglement than others. Since our techniques compress the state vector, more entanglement leads to less compressible vectors. Thus, our technique does better when there is less entanglement.

\subsection{Discussion} \label{discussion}
We successfully increase the maximum simulation size from 45 qubits to 61 qubits for the Grover's search circuit simulation on the Argonne Theta supercomputer using less than 0.8 PB. As for simulations of the other quantum applications, we have compression ratios from 4.85X to 21.34X. The results provide  evidence that we are able to simulate those applications with  47 to 49 qubits on Theta. In other words, for simulations of general circuits, our approach can increase the simulation size by 2 to 16 qubits. In this work, we seek to achieve high-fidelity simulation results for general circuits. For different purposes that do not require high-simulation fidelity, the compression ratios and simulation size would be further increased. 

We use the compression ratios to estimate our approach on the Summit supercomputer \cite{osti_1259664} at Oak Ridge National Laboratory (ORNL). The expected maximum simulation size for general circuits is 63 qubits. In 2021, we will have the exascale supercomputing system Aurora \cite{aurora2021} at Argonne. The estimated maximum simulation size will be increased from 48 qubits to 64 qubits. Our compression techniques can combine with other simulation techniques \cite{chen201864, haner20170, li2018quantum, zulehner2019accuracy, zulehner2019matrix} to scale quantum circuit simulation.

We trade time and slight fidelity for memory space. This trade-off makes the classical simulation of several quantum circuits possible, while the simulation time grows linearly with the number of gates. The time complexity is polynomial with circuit depth.

\section{Conclusion and Future Work} \label{conclusion}
Our approach performs full-state simulation of general quantum circuits using data compression techniques. This method allows the Schr\"{o}dinger-style simulation to trade time and simulation fidelity for memory space to increase the simulation size. Our compression techniques can combine with other simulation techniques \cite{chen201864, haner20170, li2018quantum, zulehner2019accuracy, zulehner2019matrix}. By using our lossy compression, we can compress state vectors and reduce memory footprints significantly compared with the existing techniques \cite{smelyanskiy2016qhipster, haner20170}. The compression errors are not correlated to the data, and hence the errors might be used to further simulate noise on real devices. The modern noise simulations add errors to perfect simulations. However, we could further adapt our lossy compression errors to noise models and then build a simulation which models noise naturally. In addition, we plan to implement our simulator on GPU-based supercomputing systems to reduce the compression and decompression time.

In summary, we have described our simulation techniques to reduce the memory requirement of full-state quantum circuit simulations by using data compression techniques. Our method provides a new option in the set of simulation tools to scale quantum circuit simulations. We propose a novel lossy compression technique to optimize compression ratios and compression speed for quantum circuit simulations. Using our approach, the memory requirement of simulating the 61-qubit Grover's search algorithm is reduced from 32 exabytes to 768 terabytes of memory on the Argonne Theta supercomputer using 4,096 nodes. We also present experimental results of random circuits, QAOA, and QFT simulations to show that our technique can achieve general circuit simulations. The compression ratio results further suggest that our technique can increase the simulation size by 2 to 16 qubits for general quantum applications. Our simulator provides a platform for quantum software debugging and quantum hardware validation.
\section*{Acknowledgments}
\footnotesize
This research was supported by the Exascale Computing Project (ECP), Project Number: 17-SC-20-SC, a collaborative effort of two DOE organizations - the Office of Science and the National Nuclear Security Administration, responsible for the planning and preparation of a capable exascale ecosystem, including software, applications, hardware, advanced system engineering and early testbed platforms, to support the nation's exascale computing imperative. This research used the resources of the Argonne Leadership Computing Facility, which is a U.S. Department of Energy (DOE) Office of Science User Facility supported under Contract DE-AC02-06CH11357. Yuri Alexeev, Hal Finkel, and Xin-Chuan Ryan Wu were supported by the DOE Office of Science. This work is also supported by the National Science Foundation under Grant No. 1619253. This work is funded in part by EPiQC, an NSF Expedition in Computing, under grants CCF-1730449/1832377, and in part by STAQ, under grant NSF Phy-1818914.

%
\bibliographystyle{ACM-Reference-Format}
\bibliography{ref}

%

\end{document}